\documentclass[final,authoryear,3p,times,10pt]{elsarticle}
\usepackage{graphicx}
\usepackage{amssymb}
\usepackage{graphicx}
\usepackage{dcolumn}
\usepackage{bm}
\usepackage{color}
\usepackage{graphicx}
\usepackage{multirow}
\usepackage{tabularx}
\usepackage{tikz}
\usepackage{pgfplots}
\usepackage{pgflibraryarrows}
\usepackage{pgflibrarysnakes}
\usepackage{threeparttable}
\usepackage{booktabs}
\usepackage{multicol}
\usepackage{longtable}
\usepackage{lipsum}
\usepackage{supertabular}

\begin{document}

\begin{frontmatter}

\title{Short term prediction of extreme returns based on the recurrence
  interval analysis} 

\author[ECUST,BU]{Zhi-Qiang Jiang}\ead{zqjiang@ecust.edu.cn}
\author[HNU,BU]{Gang-Jin Wang} \ead{wanggangjin@hnu.edu.cn}
\author[UFAL,BU]{Askery Canabarro} \ead{askery@gmail.com}
\author[UR]{Boris Podobnik} \ead{bp@phy.hr}
\author[HNU]{Chi Xie} \ead{xiechi@hnu.edu.cn}
\author[BU]{H. Eugene Stanley} \ead{hes@bu.edu}
\author[ECUST]{Wei-Xing Zhou \corref{cor1}}
\cortext[cor1]{Corresponding author. Address: 130 Meilong Road, P.O. Box
  114, School of Business, 
              East China University of Science and Technology, Shanghai
              200237, China, Phone: +86 21 64253634, Fax: +86 21
              64253152.} 
\ead{wxzhou@ecust.edu.cn} 
\address[ECUST]{Department of Finance, East China University of Science
  and Technology, Shanghai 200237, China} 
\address[HNU]{Business School and Center of Finance and Investment
  Management, Hunan University, Changsha 410082, 
China}
\address[BU]{Boston University, Boston, MA 02215, USA}
\address[UFAL]{Universidade Federal de Alagoas, 57309-005, Arapiraca-AL, Brazil}
\address[UR]{Zagreb School Economics and Management, 41000 Zagreb, Croatia}

\begin{abstract}

\noindent
Being able to predict the occurrence of extreme returns is important in
financial risk management. Using the distribution of recurrence
intervals---the waiting time between consecutive extremes---we show
that these extreme returns are predictable on the short term. Examining
a range of different types of returns and thresholds we find that
recurrence intervals follow a $q$-exponential distribution, which we
then use to theoretically derive the hazard probability $W(\Delta t
|t)$. Maximizing the usefulness of extreme forecasts to define an
optimized hazard threshold, we indicates a
financial extreme occurring within the next day when the hazard
probability is greater than the optimized threshold. Both
in-sample tests and out-of-sample predictions indicate that these
forecasts are more accurate than a benchmark that ignores the predictive
signals. This recurrence interval finding deepens our understanding of
reoccurring extreme returns and can be applied to forecast extremes in risk management.


\end{abstract}


\end{frontmatter}


\section{Introduction}

\noindent
Predicting such extreme financial events as market crashes, bank
failures, and currency crises is of great importance to investors and
policy markers because they destabilize the financial system and can
greatly shrink asset value. Much research has been carried out in an
attempt to detect the underlying vulnerabilities and the common
precursors to financial extremes. A number of different models have been
developed to predict the occurrence of financial distresses including those
using probability \citep{Martin-1977-JBF, Canbas-Cabuk-Kilic-2005-EJOR,
  Barrell-Davis-Karim-Liadze-2010-JBF, Tinoco-Wilson-2013-IRFA,
  Li-Wang-2014-KBB, Laina-Nyholm-Sarlin-2015-RFE}, signal approaches
\citep{Kaminsky-Lizondo-Reinhart-1998-SPIMF, Edison-2003-IJFE,
  Duan-Claustre-2008-CER, Christensen-Li-2014-JFS} and intelligence
\citep{Kumar-Ravi-2007-EJOR, Demyanyk-Hasan-2010-Omega}.  A
faster-than-exponential increase in price accompanied by accelerating
price oscillations indicates the presence of bubbles
\citep{Sornette-2003, Sornette-Cauwels-2015-RBE}. The behavior of these
bubbles can be characterized using the log-period power-law singularity
(LPPLS) model, which is capable of accurately forecasting a bubble's
tipping point \citep{Sornette-Woordard-Zhou-2009-PA,
  Jiang-Zhou-Sornette-Woodard-Bastiaensen-Cauwels-2010-JEBO,
  Sornette-Demos-Zhang-Cauwels-Filimonov-Zhang-2015-JIS}.

Recent research on the occurrence of financial extremes and on the
market dynamics around financial crashes has enabled us to better
forecast emerging financial crises. We can understand the occurrence
pattern of extremes by determining the distribution of waiting times
between consecutive financial extremes (the ``recurrence intervals'')
and charting the memory behavior within the occurring extremes.
\cite{Bogachev-Bunde-2009-PRE,
  Jiang-Canabarro-Podobnik-Stanley-Zhou-2016-QF} built an early warning
model of this waiting time distribution to predict the probability that
extremes will occur within a given time period. Following a financial
crisis the financial system gradually transitions back to a stasis
\citep{Bussiere-Fratzscher-2006-JIMF}. This relaxation behavior
following a financial market crash is similar to the aftershocks
following an earthquake \citep{Lillo-Mantegna-2003-PRE,
  Petersen-Wang-Havlin-Stanley-2010-PRE}.  \cite{Sornette-2003}
indicates that a possible theoretical explanation for bursts of
speculating bubbles is a positive herding behavior of traders that
causes local self-excited crashes
\citep{Gresnigt-Kole-Franses-2015-JBF}. This is in accordance with the
phenomenon that extremes cluster and are
interdependent. \cite{Gresnigt-Kole-Franses-2015-JBF} show that
approximately 76--85\% of occurring extremes are triggered by other
extremes, and they develop an early warning model that treats financial
crashes as earthquakes and compute the probability that an extreme event
will occur within a certain time period.

Here we extend the probabilistic framework for extreme returns
presented in \cite{Jiang-Canabarro-Podobnik-Stanley-Zhou-2016-QF} to
predict extremes by using the conditional probability of an future
extreme event within a fixed time frame in which Type 1 and Type 2
errors are balanced in current market state. The contributions of our works are in four ways.  

\begin{itemize}

\item[{(i)}] We identify extremes by locating the threshold at the
  minimum KS value between the empirical and fitting distributions of
  the extreme values.

\item[{(ii)}] We classify the returns as either extreme or non-extreme
  by quantifying the extreme threshold, and we assume that the extremes
  are independent. This simplifies the modeling and reduces the
  computational complexity when estimating parameters but provides an
  adequate performance when doing out-of-sample prediction.

\item[{(iii)}] We define a hazard probability that is dependent on the
  distribution formula of recurrence intervals between extremes, and
  this translates the problem into finding a suitable distribution form
  for recurrence intervals. Unlike the Hawkes point process, our
  modeling framework is easy to implement.  

\item[{(iv)}] Instead of using a predefined threshold of hazard
  probability, we predict extremes when the hazard probability exceeds an optimized hazard threshold, 
  obtained by maximizing a usefulness function
  that takes into account an investor's preference for either Type 1 or
  Type 2 errors.

\end{itemize}

We organize the paper as follows. In Section 2 we present a brief review
of recurrence interval analysis and early warning models. In Section 3
we provide the dataset. In Section 4 we describe the Model and
Methods. In Section 5 we present the results of our recurrence interval
analysis for different subperiods. In Section 6 we document and discuss
the performance of our out-of-sample predictions. In Section 7 we
present our conclusions.

\section{Literature  review}

\subsection{Recurrence intervals analysis}

\noindent
Recurrence intervals, defined as the time periods between consecutive
extreme events, have been a topic of extensive research across many
fields, financial markets in particular. The primary contribution of the
published research is an understanding of the statistical regularities
in recurrence intervals. The memory behavior in the underlying process
strongly affects the distribution form of recurrence intervals
\citep{Chicheportiche-Chakraborti-2013-XXX,Chicheportiche-Chakraborti-2014-PRE}. 
The interval distribution is exponential if the process has no
memory. Incorporating a long memory into the underlying process greatly
alters the recurrence interval distribution. For example, the stretched
exponential and Weibull recurrence interval distribution are
analytically and numerically confirmed in a process with a long linear
memory \citep{Santhanam-Kantz-2008-PRE}. When a process has a long
nonlinear memory (a multifractual process), the recurrence intervals are
power-law distributed \citep{Bogachev-Eichner-Bunde-2007-PRL}.

There is extensive literature that examines the empirical distribution
of recurrence intervals in financial markets. The distribution form is
found to be dependent on data source, data type, and data
resolution. For example, recurrence interval distributions with a
power-law tail are found in the daily volatilities in the Japanese
market \citep{Yamasaki-Muchnik-Havlin-Bunde-Stanley-2005-PNAS}, in the
minute volatilities in the Korean \citep{Lee-Lee-Rikvold-2006-JKPS} and
Italian markets \citep{Greco-SorrisoValvo-Carbone-Cidone-2008-PA}, in
the daily returns in the US stock markets
\citep{Bogachev-Eichner-Bunde-2007-PRL, Bogachev-Bunde-2009-PRE}, in the
minute returns in the Chinese markets \citep{Ren-Zhou-2010-NJP}, and in the
minute volume in the US \citep{Li-Wang-Havlin-Stanley-2011-PRE} and
Chinese markets \citep{Ren-Zhou-2010-PRE}. In addition, stretched
recurrence interval distributions are also observed in the financial
volatility at different resolutions in a range of different markets
\citep{Wang-Wang-2012-CIE,Xie-Jiang-Zhou-2014-EM,
  Jiang-Canabarro-Podobnik-Stanley-Zhou-2016-QF}. The $q$-exponential
distribution has also been observed in the recurrence intervals between
losses in financial returns
\citep{Ludescher-Tsallis-Bunde-2011-EPL,Ludescher-Bunde-2014-PRE}, and
the corresponding distribution in the Chinese stock index future market
is a stretched exponential \citep{Suo-Wang-Li-2015-EM}.

In addition to the inconsistent findings on the distribution of
empirical recurrence intervals, the existence of scaling behaviors in
the recurrence interval distribution for the extremes filtered by
different thresholds is under debate. Analyzing the distribution of
recurrence intervals has indicated that the extreme event filtering
threshold should influence the recurrence interval distribution
\citep{Xie-Jiang-Zhou-2014-EM, Chicheportiche-Chakraborti-2014-PRE,
  Suo-Wang-Li-2015-EM,
  Jiang-Canabarro-Podobnik-Stanley-Zhou-2016-QF}. This indication was
supported when the estimated distributional parameters were found to be
strongly dependent on the thresholds when the recurrence intervals are
fitted by such distribution functions as the stretched exponential
distribution \citep{Xie-Jiang-Zhou-2014-EM, Suo-Wang-Li-2015-EM,
  Jiang-Canabarro-Podobnik-Stanley-Zhou-2016-QF} and the $q$-exponential
distribution \citep{Ludescher-Tsallis-Bunde-2011-EPL,
  Chicheportiche-Chakraborti-2014-PRE,
  Jiang-Canabarro-Podobnik-Stanley-Zhou-2016-QF}.
\cite{Ludescher-Tsallis-Bunde-2011-EPL} and
\cite{Ludescher-Bunde-2014-PRE} propose that the distribution of
recurrence intervals depends only on the mean recurrence interval
$\tau_Q$, and not on a specific asset or on the time resolution of the
data.

Only a limited amount of research has used recurrence interval analysis
to assess and manage risks in financial markets. An improved method for
estimating the value at risk (VaR) based on the recurrence interval is
significantly more accurate than traditional estimates based on the
overall or local return distributions
\citep{Bogachev-Bunde-2009-PRE,Ludescher-Tsallis-Bunde-2011-EPL}. Another
way of predicting extremes using statistics of recurrence intervals is
also superior to the precursory pattern recognition technique when the
underlying process is multifractal
\citep{Bogachev-Bunde-2009-PRE}. Defining a conditional loss probability
as the inverse of the expected waiting time before observing another
extreme determined by the latest recurrence interval,
\cite{Ren-Zhou-2010-NJP} finds that the risk of extreme loss events is
high if the latest recurrence interval is long or short. In all of these
studies, however, only in-sample tests are conducted, and a good
performance in in-sample tests cannot ensure good results in
out-of-sample tests. In contrast,
\cite{Jiang-Canabarro-Podobnik-Stanley-Zhou-2016-QF} recently found that
the extreme predicting method using recurrence interval analysis does
provide good predictions in out-of-sample tests.

\subsection{Early warning model of financial crisis}

Such events as market crashes, currency crises, and bank failures are
financial crisis in which the value of assets or the equity of financial
institutions shrinks rapidly. Financial crises shock the real-world
economy and can cause recessions or depressions if left unchecked. To
reduce investor losses and shocks to the economy and to reduce financial
turbulence, much effort has gone into predicting financial
extremes. There is a plethora of literature on forecasting financial
crises, especially currency crises and bank failures, and most of the
research relies on the early warning model (EWM)
\citep{Kumar-Ravi-2007-EJOR, Demyanyk-Hasan-2010-Omega}. The EWM
identifies the leading indicators of emerging financial problems and
uses such techniques as logit (or probit) regressions and intelligence
approaches to translate them into the hazard probability of crises
occurring in the future, which is used as an early warning signal that
indicates whether a crisis is imminent.

Compared to the vast EWM research predicting bank failures and currency
crisis, early warning models to monitor stock markets and provide
warning signals of market extremes have received little attention. The
contributions of the existing literature are as follows.

A number of indicators are able to warn of incoming financial
extremes. \cite{Coudert-Gex-2008-JEF} show that risk aversion indicators
are useful in predicting stock market crises, but not currency
crises. \cite{Chen-2009-JBF} finds that such macroeconomic indicators as
yield curve spreads and inflation rates can be used to predict stock
market recessions. \cite{Alessi-Detken-2011-EJPE} show that a global
measure of liquidity can predict asset price
booms. \cite{Herwartz-Kholodilin-2014-JFo} show that the price-to-book
ratio can predict emerging price
bubbles. \cite{Li-Chen-French-2015-QREF} show that such variables of
index futures and options as the VIX, open interest, dollar volume, put
option price, and put option effective spread can predict equity market
crises. \cite{Chang-Hu-Kao-Chang-2015-AE} define the average value at
risk (AVaRs) based on the ARMA-GARCH model with standard infinitely
divisible innovations as an early warning indicator and find that AVARs
can predict both extreme events and highly volatile markets. By
constructing two investment networks based on the cross-border equity
and a long-term debt securities portfolio,
\cite{Joseph-Joseph-Chen-2014-SR} identify two network-based indicators
(algebraic connectivity and edge density) that could have predicted the
2008 global financial crisis.
\cite{Minoiu-Kang-Subrahmanian-Berea-2015-QF} show that the
interconnectedness in the global network of financial linkages could
have predicted the financial crises that occurred during the 1978--2010
period.

Composite indices averaged from crisis-related variables have been
proposed to predict financial crises. \cite{Oh-Kim-Kim-2006-ES} propose
a daily financial condition indicator, market volatility, to determine
whether a stock market is unstable or not. \cite{Kim-Lee-Oh-Kim-2009-ES}
define and propose a stock market instability index based on the
difference between the current market condition and the past conditions
when the market was stable. \cite{Son-Oh-Kim-Kim-2009-ESA} propose a
model to predict stock market collapse that signals when a massive
selling by global institutional investors occurs.
\cite{Ahn-Oh-Kim-Kim-2011-ESA} integrate all crisis-related variables
into a monthly financial market condition indicator and find that by
using a support vector machine the indicator can detect market
crises. \cite{Yoon-Park-2014-DSP} use a market instability index to
capture risk warning levels, quantify the instability level of the
current market, and predict its future behavior.

There is a pattern of price trajectories that signals near-future market
crashes. \cite{Sornette-2003} develops a log-periodic power law
singularity (LPPLS) model for detecting bubbles by combining (i) the
economic theory of rational expectation bubbles, (ii) the effect on the
market of imitation and herding behaviors among investors and traders,
and (iii) the mathematical and statistical physics of bifurcations and
phase transitions. The faster-than-exponential (power law with
finite-time singularity) increase in asset prices accompanied by
accelerating oscillations is the main diagnostic that indicates bubbles
\citep{Sornette-Woordard-Zhou-2009-PA,
  Jiang-Zhou-Sornette-Woodard-Bastiaensen-Cauwels-2010-JEBO,
  Sornette-Demos-Zhang-Cauwels-Filimonov-Zhang-2015-JIS}. \cite{KurzKim-2012-AEL}
also corroborate that the LPPLS pattern can be used as an early warning
signal for market crashes. In addition,
\cite{Yan-vanTuyllvanSerooskerken-2015-PLoS1} convert the price series
into networks using a visible graph alogorithm and use the
degree-of-price network to measure the magnitude of the
faster-than-exponential growth of stock prices, and to predict imminent
financial extreme events. On average this indicator performs better than
the LPPLS pattern-recognition indicator.

The patterns of financial crises are modeled to predict financial
extreme events. \cite{Jiang-Canabarro-Podobnik-Stanley-Zhou-2016-QF}
uncover the distribution pattern of waiting time between consecutive
market extremes and use it to define a hazard probability that
subsequent extremes will occur within a certain time period. They find
that this hazard probability performs well in out-of-sample
predictions. As an analogue to the seismic activity around earthquakes,
\cite{Gresnigt-Kole-Franses-2015-JBF} adopt an epidemic-type aftershock
sequence model (a type of mutually self-exciting Hawkes point process)
to capture the occurring dynamics of stock market crashes, which can
serve as an early warning model for predicting the probability of
medium-term crashes.

\section{Data sets}

\noindent
We analyze the daily Dow Jones Industrial Average (DJIA) index from 16
February 1885 to 31 December 2015. The logarithmic return of the DJIA
index over a time scale of one day is defined
\begin{equation}
  r(t) = \ln I(t) - \ln I(t-1).
  \label{Eq:Return}
\end{equation}

\begin{figure}[htbp]
 \centering
  \includegraphics[width=7.2cm]{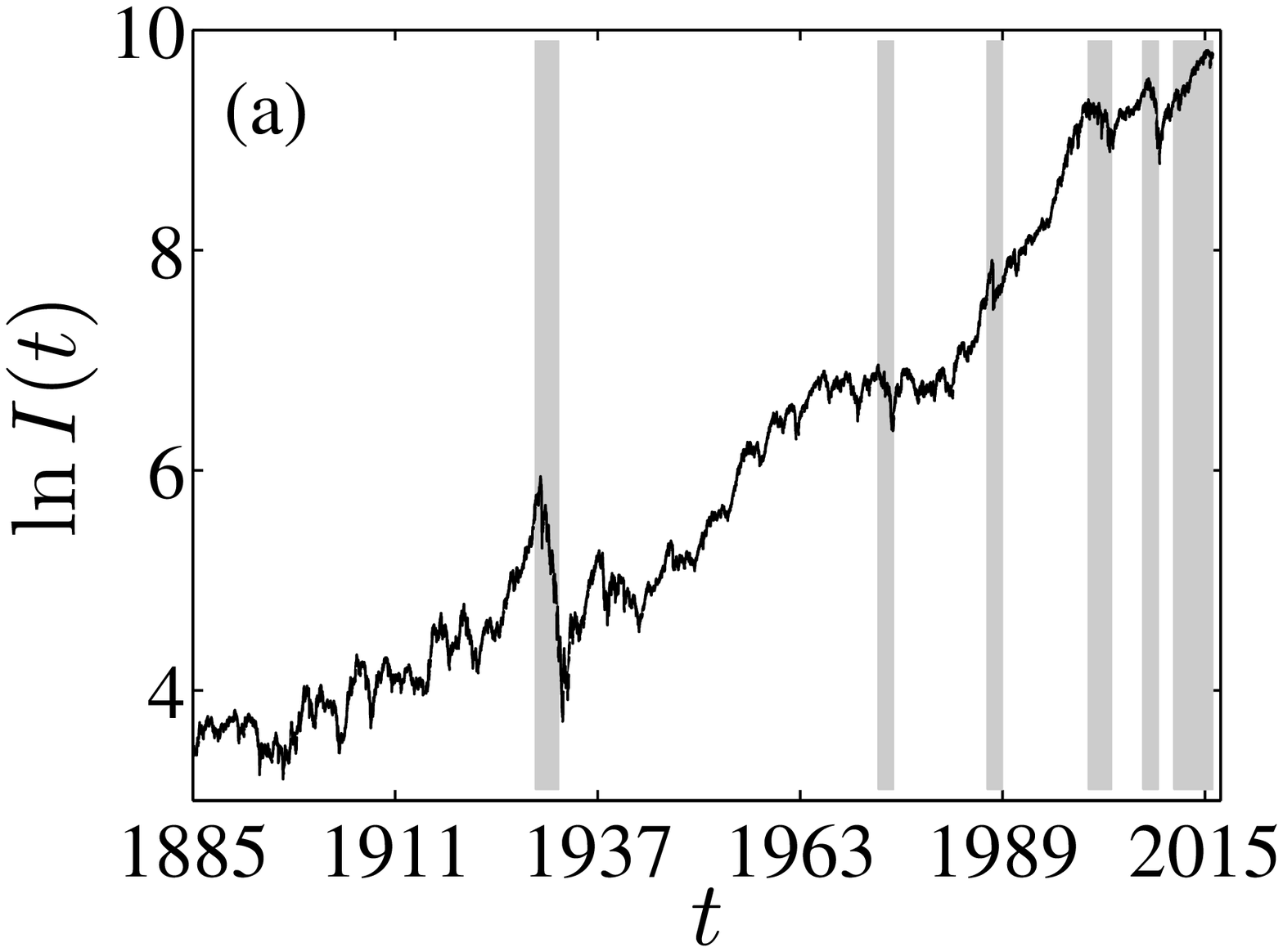}
  \includegraphics[width=7.5cm]{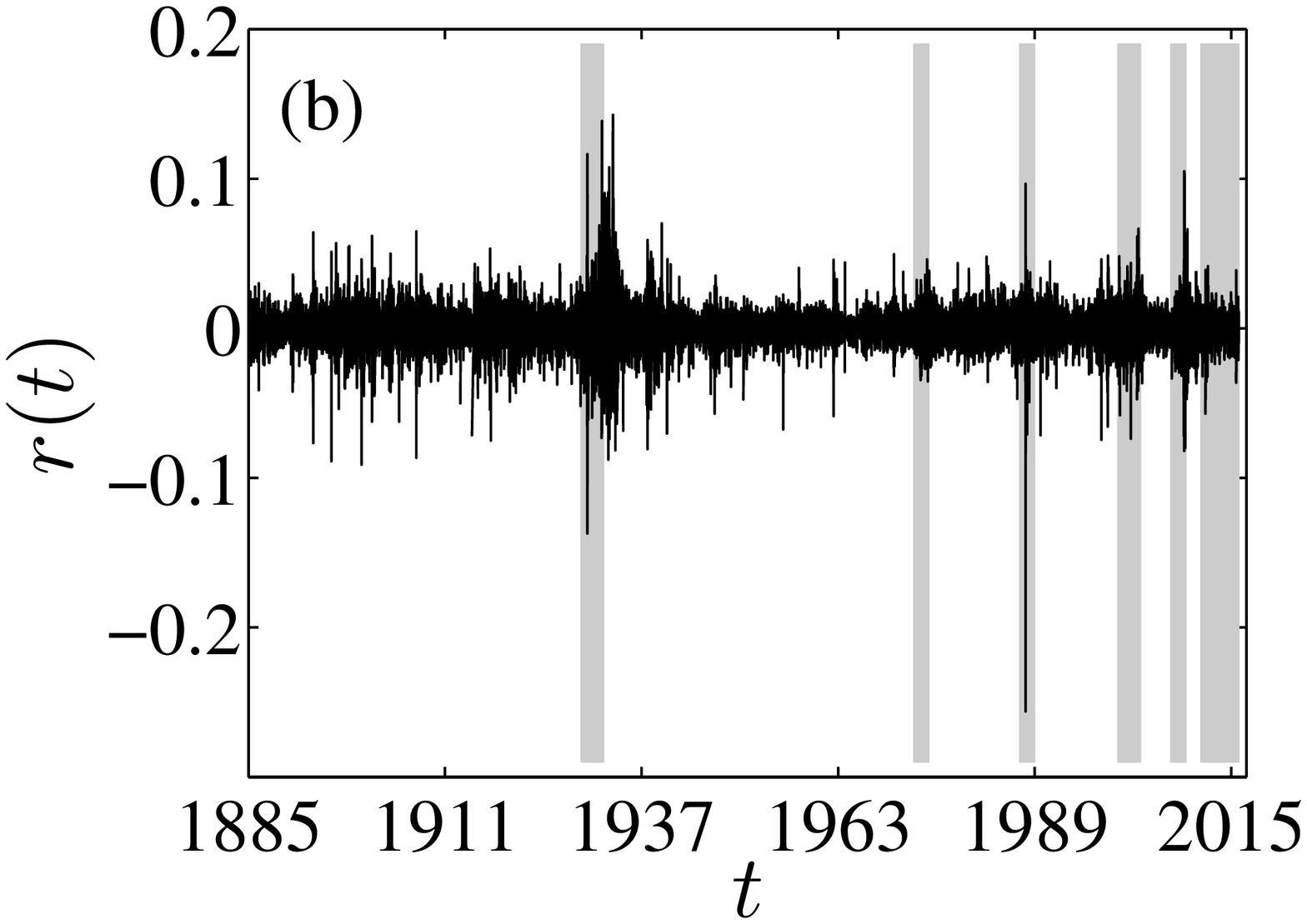}
 \caption{\label{Fig:Index:Return} (color online). Plots of the
   logarithmic DJIA index $\ln I(t)$ and it's difference, return
   $r(t)$. (a) $\ln I(t)$. (b) $r(t)$. }
\end{figure}

Figures~\ref{Fig:Index:Return}(a) and \ref{Fig:Index:Return}(b) show
plots of the logarithmic DJIA and its return, respectively. The DJIA
index grows from 30.92 on 16 February 1885 to 17425.03 on 31 December
2015 with a total logarithmic return greater than 6. Although the index
exhibits a rising trend throughout sample period, there are falling
trends and range-bounds in different
subperiods. Figure~\ref{Fig:Index:Return} shows six turbulent periods
(highlighted in shadow), the Wall Street crash of 1929--1932, the oil
crisis of 1973--1975, the Black Monday crash of 1987--1989, the dot-com
bubble of 2000--2003, the subprime crisis 2007--2009, the 2008 financial
crisis, and the European sovereign debt crisis 2011--2015.

\section{Model and Methods}

\subsection{Identifying extreme returns}
\label{Sec:Method:EVT}

\noindent
An extreme value is usually defined as a peak above a threshold (POT)
\citep{Ren-Zhou-2010-PRE, Alessi-Detken-2011-EJPE,
  Christensen-Li-2014-JFS, Sevim-Oztekin-Bali-Gumus-Guresen-2014-EJOR,
  Suo-Wang-Li-2015-EM} that is $m$ times the sample standard
deviation. The parameter $m$ is a predefined value (see a summary in
Table 1 of \cite{Sevim-Oztekin-Bali-Gumus-Guresen-2014-EJOR}). Although
identifying extreme events in terms of POT is widely applied in
empirical analysis, the POT has drawbacks. A small $m$ value will
produce many ``extreme values,'' not all of which are truly extreme, and
a large $m$ value will indicate genuine extremes but not necessarily
include all of them.

According to extreme value theory, the distribution of extreme values
differs from that of non-extreme values. Finding the extreme values is
equivalent to finding a group of data ($x \ge x_t$) that satisfies the
extreme value distribution \citep{Cumperayot-Kouwenberg-2013-JIMF}
\begin{equation}\label{Eq:ExtDis}
G(x) = \left\{ \begin{array}{ll}
\exp\left[ -(1+\gamma\frac{x-\mu}{\sigma}) ^{-1/\gamma}  \right] &
\textrm{for~~~~ $\gamma \neq 0$}  \\ 
\exp\left[- \exp(-\frac{x-\mu}{\sigma}) \right] & \textrm{for~~~~ $\gamma = 0$}
\end{array} \right.,
\end{equation}
where $G(x)$ is the cumulative distribution of the generalized extreme
value distribution, and $\mu$, $\sigma$, and $\gamma$ are location,
scale, and shape parameters, respectively, and $x_t$ is the extreme
value threshold. The inverse of the shape parameter $1/\gamma$ is simply
the tail exponent of the sample distribution.

\begin{figure}[htbp]
 \centering
  \includegraphics[width=7.3cm]{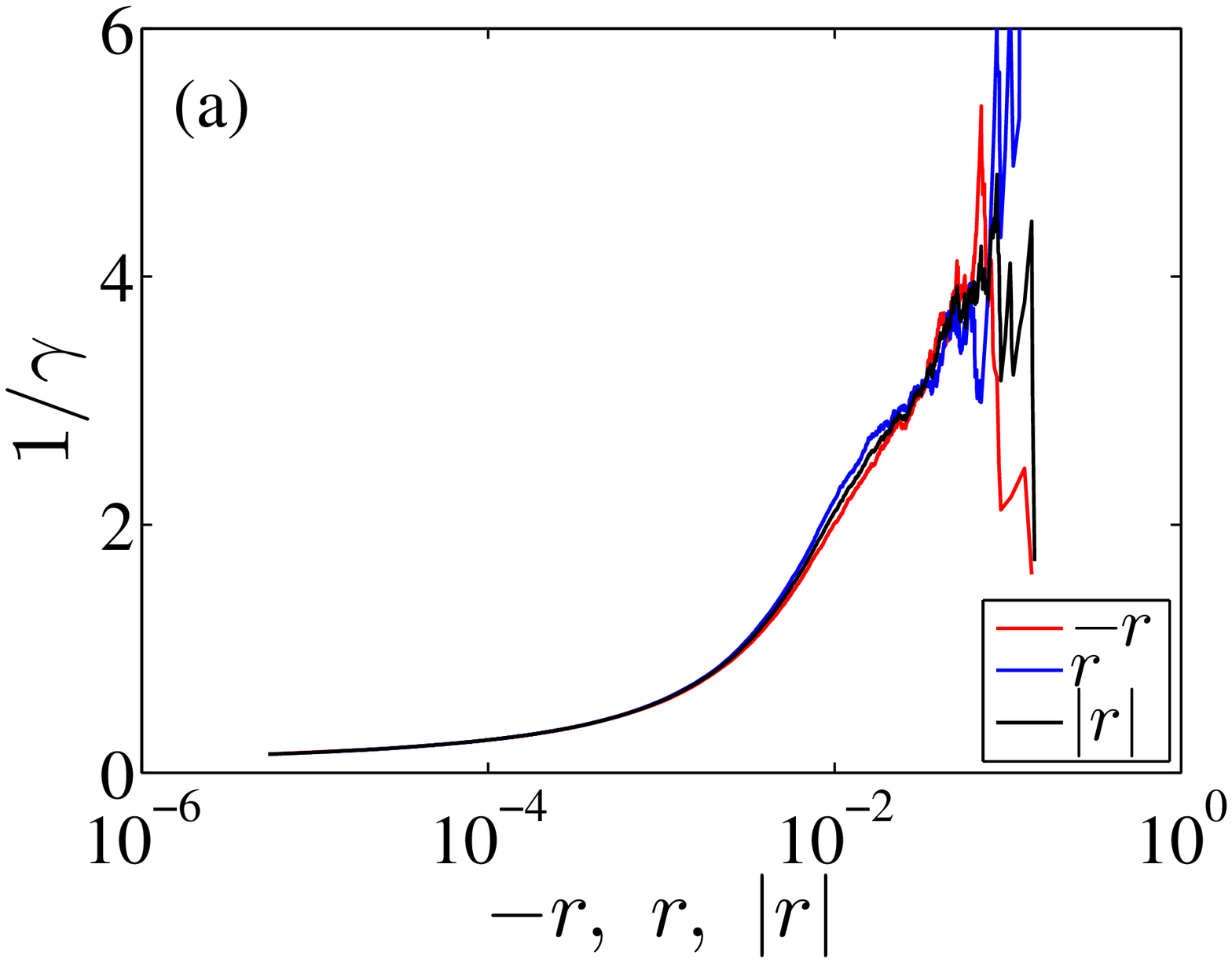}
  \includegraphics[width=7.5cm]{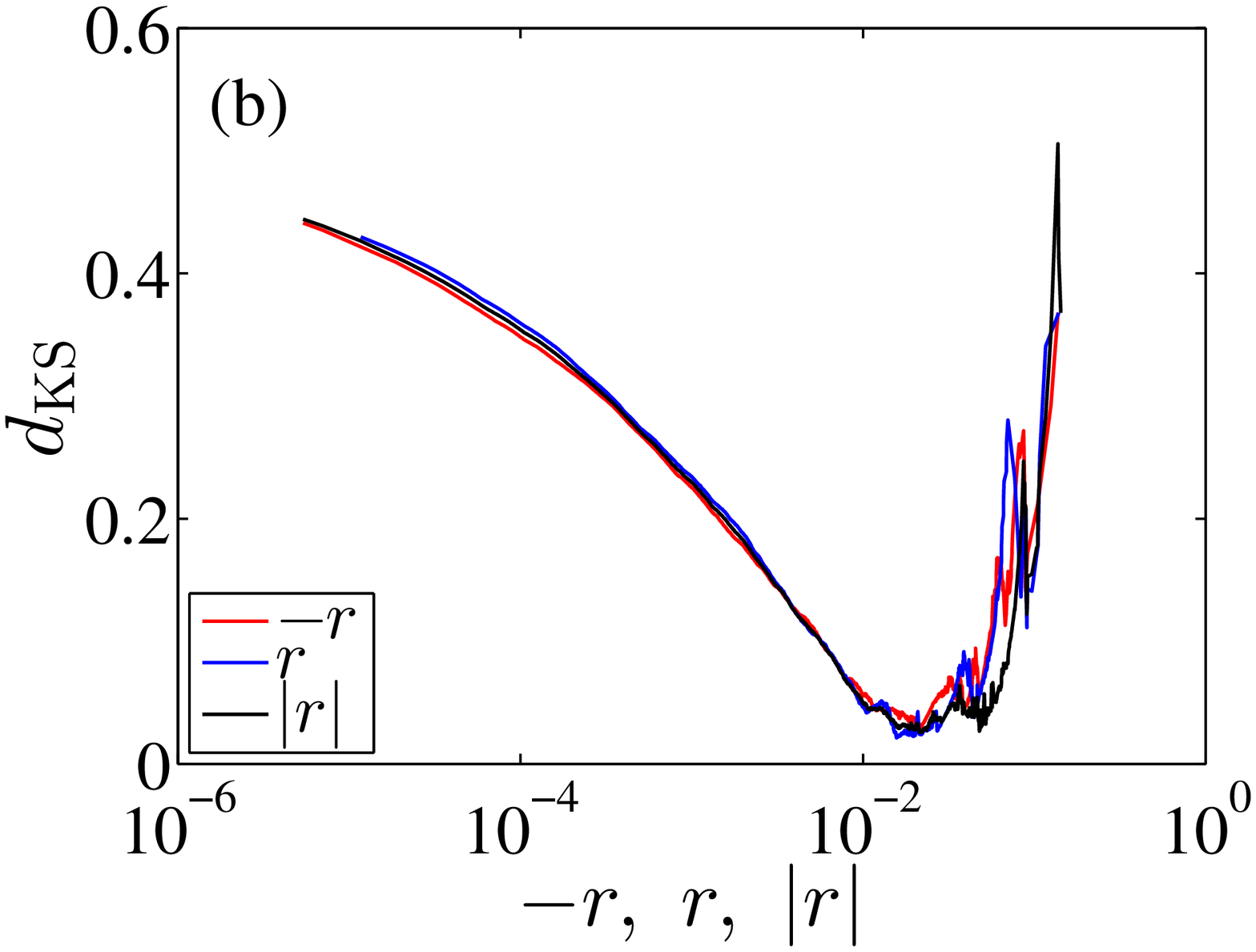}
 \caption{\label{Fig:EVT:Threshold} (color online). Determining the
   extreme value threshold $x_t$ for negative, positive, and absolute
   returns.  (a) Plots of the tail exponents $1/\gamma$ as a function of
   the sorted returns. (b) Plots of the KS statistics $d_{\rm{KS}}$ with
   respect to the sorted returns. The KS statistics is defined as the
   maximum absolute difference between the empirical and fitting tail
   distributions.}
\end{figure}

We estimate the shape parameter $\gamma$ using the Hill estimator
\citep{Hill-1975-AS}, which is a non-parametric method. For a given
sample $\{x_1, x_2, \cdots, x_n\}$, we sort the data in ascending order,
\begin{equation}\label{Eq:Sort:Data}
x_{(1)} \le x_{(2)} \le \cdots \le x_{(n)}.
\end{equation}
The $\gamma$ value given by the Hill estimator is
\begin{equation}\label{Eq:Hill}
\gamma = \frac{1}{k} \sum_{i=1}^{k} \left[\log x_{(n+1-k)} -\log x_{(k)}
  \right], 
\end{equation}
where $x_k$ corresponds to the extreme value threshold $x_t$ that will
be determined.  

One way to find threshold $x_t$ is by (i) estimating the value of
$\gamma$ with respect to all possible values of $x_t$, and (ii) plotting
$1/\gamma$ against $x_t$ to find a range of $x_t$ values within which
the estimated $1/\gamma$ values are stable
\citep{Pozo-AmuedoDorantes-2003-JIMF,
  Reboredo-ReveraCastro-MachadodeAssis-2014-QF}. In practice, this
``stable behavior'' between $1/\gamma$ and $x_k$ is difficult to
quantify. For example, Fig.~\ref{Fig:EVT:Threshold}(a) uses DJIA returns
to illustrate the estimated $1/\gamma$ as a function of the sorted DJIA
(negative, positive, and absolute) returns. The $1/\gamma$ values
strongly fluctuate and there is no stable range. An alternative approach
is to use KS statistics to measure the agreement between the empirical
and fitting tail distributions.  KS statistics quantify the maximum
absolute difference between both distributions. The most suitable
threshold $x_t$ is associated with the best fits to the tail
distribution, which has the smallest KS statistical values
\citep{Clauset-Shalizi-Newman-2009-SIAMR,
  Jiang-Xie-Li-Podobnik-Zhou-Stanley-2013-PNAS}. Figure~\ref{Fig:EVT:Threshold}(b)
shows the plots of the KS statistics $d_{\rm{KS}}$ with respect to the
sorted (negative, positive, and absolute) returns. The significant low
point in each curve allows us to more easily determine the extreme value
threshold $x_t$.

For sake of comparison, we also use the quantiles of 95\%, 97.5\%, and
99\% to define the extremes. Definitions based on the quantile are
common in the analysis of value-at-risk
(VaR). \cite{Gresnigt-Kole-Franses-2015-JBF} also define the 95\%
quantile of returns and the 95\% quantile of negative returns as
extremes and crashes.

\subsection{Determining hazard probability}

\noindent
By taking into consideration only the time in which extremes occur, we
base our prediction of extreme returns on the hazard probability
$W(\Delta t | t)$, which measures the probability that following an
extreme return occurring at $t$ time in the past there is an additional
waiting time $\Delta t$ before another extreme return occurs.
\cite{Sornette-Knopoff-1997-BSSA} and
\cite{Bogachev-Eichner-Bunde-2007-PRL} theoretically derived the hazard
probability $W(\Delta{t},t)$ using the distribution of recurrence
intervals between extreme events,
\begin{equation}
 W(\Delta{t}|t)=\frac{\int_t^{t+\Delta{t}}p(\tau){\rm{d}}
   \tau}{\int_t^{\infty}p(\tau){\rm{d}} \tau},  
 \label{Eq:Wq}
\end{equation}
where $p(\tau)$ is the probability distribution of the recurring
intervals. Once we have the distribution form of $p(\tau)$, the
formula for $W(\Delta{t}|t)$ can be derived from Eq.~(\ref{Eq:Wq}).

Although the recurrence intervals of Poisson processes are exponentially
distributed \citep{Yamasaki-Muchnik-Havlin-Bunde-Stanley-2005-PNAS,
  Bogachev-Eichner-Bunde-2007-PRL, Chicheportiche-Chakraborti-2014-PRE},
which generates a constant hazard probability when $\Delta t$ is given,
financial processes always exhibit such non-Poissonian characteristics
as long-term dependence and multifractality in volatilities
\citep{Calvet-Fisher-2002-RES}, medium-term dependence (e.g., momentum
and contrarian behaviors \citep{Chan-Jegadeesh-Lakonishok-1996-JF,
  Shi-Jiang-Zhou-2015-PLoS1}), and multiscaling behaviors in returns
\citep{Calvet-Fisher-2002-RES}, which leads to that the recurrence intervals are no longer exponentially
distributed, and that the derivation of the close distribution form for
the recurrence intervals is obstructed
\citep{Chicheportiche-Chakraborti-2013-XXX}. The non-Poissonian features also result in a controversial situation in the
empirical analysis of the distribution formula of recurrence
intervals. For example, the reported distributions range from a power-law
distribution with an exponential cutoff
\citep{Yamasaki-Muchnik-Havlin-Bunde-Stanley-2005-PNAS,
  Lee-Lee-Rikvold-2006-JKPS, Greco-SorrisoValvo-Carbone-Cidone-2008-PA,
  Ren-Zhou-2010-NJP} to a stretched exponential distribution
\citep{Wang-Wang-2012-CIE, Suo-Wang-Li-2015-EM,
  Jiang-Canabarro-Podobnik-Stanley-Zhou-2016-QF}, from a $q$-exponential
distribution \citep{Ludescher-Tsallis-Bunde-2011-EPL,
  Ludescher-Bunde-2014-PRE, Chicheportiche-Chakraborti-2014-PRE} to a
$q$-Weibull distribution
\citep{Reboredo-ReveraCastro-MachadodeAssis-2014-QF}. Here we employ
three common functions to fit the recurrence interval distributions. The
three formulas are the stretched exponential distribution,
\begin{equation}\label{Eq:PDF:sExp}
p(\tau) = a  \exp \left[-(b\tau)^\mu \right],
\end{equation}
the $q$-exponential distribution,
\begin{equation}\label{Eq:PDF:qExp}
p(\tau) = (2-q) \lambda [1+(q-1)\lambda \tau]^{-\frac{1}{q-1}},
\end{equation}
and the Weibull distribution,
\begin{equation}\label{Eq:PDF:WBL}
p(\tau) =  \frac{\alpha}{\beta}
\left(\frac{\tau}{\beta}\right)^{\alpha-1} \exp
\left[-\left(\frac{\tau}{\beta}\right)^\alpha \right], 
\end{equation}

By putting the three probability distributions
Eqs.~(\ref{Eq:PDF:sExp})--(\ref{Eq:PDF:WBL}) into Eq.~(\ref{Eq:Wq}), we
obtain the hazard probability $W_{\rm{sE}}$ for the stretched
exponential distribution,
\begin{equation}\label{Eq:sExp:Hazard}
W_{\rm{sE}}(\Delta t |t) = \frac{\frac{b\mu}{a} -
  \Gamma_l\left(\frac{1}{\mu}, (bt)^{\mu}\right) -
  \Gamma_u\left(\frac{1}{\mu}, [b(t+\Delta
    t)]^\mu\right)}{\Gamma_u\left( \frac{1}{\mu}, (bt)^{\mu}\right)}, 
\end{equation}
the hazard probability $W_{\rm{qE}} (\Delta t | t)$ for $q$-exponential
distribution, 
\begin{equation}\label{Eq:qExp:Hazard}
W_{\rm{qE}}(\Delta t |t) = 1 - \left[ 1 + \frac{(q-1)\lambda \Delta t}{1
    + (q-1) \lambda t} \right] ^{1-\frac{1}{q-1}}, 
\end{equation}
and the hazard probability $W_{\rm{W}}(\Delta t |t)$ for Weibull
distribution, 
\begin{equation}\label{Eq:WBL:Hazard}
W_{\rm{W}}(\Delta t |t) = 1-\exp \left[
  \left(\frac{t}{\beta}\right)^\alpha  - \left(\frac{t + \Delta
    t}{\beta}\right)^\alpha \right], 
\end{equation}
where $\Gamma_l(s,x)$ and $\Gamma_u(s,x)$ are lower and upper incomplete
gamma functions. For fixed $\Delta t$, all three hazard probabilities
decrease as $t$ increases, which explains the clustering of extremes in
financial returns and volatilities.

To use the hazard probability $W(\Delta t|t)$ to predict the extremes we
must set a hazard threshold $w_t$ to trigger the early warning indicator
of an approaching extreme event. If the hazard probability $W(\Delta
t|t)$ is greater than the hazard threshold $w_t$, an alarm that an
extreme return will occur during the next $\Delta t$ time is
activated. The hazard threshold $w_t$ is not an arbitrary given value
but---depending on the risk level preferences of investors---is
optimized to balance between false alarms and not detecting events.

\subsection{Evaluating predicting signals}

\noindent
The hazard probability $W(\Delta t | t)$
becomes a binary extreme forecast that equals one when $W(\Delta t | t)$
exceeds the hazard threshold $w_t$ and equals zero otherwise. When
comparing the forecasted extremes with the actual events we see (i)
correct predictions of an extreme return occurring, (ii) correct
predictions of a non-extreme return occurring, (iii) missed events, and
(iv) false alarms. By counting how many times each outcome occurs we can
compute a range of evaluation measurements including the correct
prediction rate, the false alarm rate, and the accuracy. Our primary
interest here is correct prediction rate $D$ and false alarm rate $A$,
which are defined as
\begin{equation}
 D = \frac{n_{11}}{ n_{01} + n_{11}},~~A = \frac{n_{10}}{n_{00} +
   n_{10}}, 
 \label{Eq:Wq:AD}
\end{equation}
where $n_{11}$ is the number of extreme returns that are correctly
predicted, $n_{00}$ the number of non-extreme returns that are correctly
predicted, $n_{01}$ the number of missed events, and $n_{10}$ the number
of false alarms. Following \cite{Gresnigt-Kole-Franses-2015-JBF}, we use
the Hanssen-Kuiper skill score (KSS) to assess the validity of extreme
forecasts. The KSS is the difference $D-A$ between the correct
prediction rate and the false alarm rate.  The KSS encompasses both
missing occurrence errors and false alarms errors. Decreasing these two
errors increases the value of KSS.

Our goal is to find a balanced signal for investors when they prefer
either Type 1 and Type 2 errors and to take into account whether they
use or discard the predictive signals. Following
\cite{Alessi-Detken-2011-EJPE} we define a loss function when a hazard
probability threshold is added issue extreme forecasts,
\begin{equation}
 L(\theta) = \theta (1-D) + (1-\theta) A,
 \label{Eq:Wq:Loss}
\end{equation}
where $1 - D$ is the ratio of missing events (Type 1 errors) and $A$ is
the ratio of false alarms (Type 2 errors). The parameter $\theta$ is the
investor preference for avoiding either Type 1 or Type 2 errors
\citep{ElShagi-Knedlik-vonSchweinitz-2013-JIMF}.

We further define the usefulness of extreme forecasts as
\begin{equation}
 U(\theta) = \min( \theta, 1-\theta) - L(\theta),
 \label{Eq:Wq:Usefulness}
\end{equation}

where $\min(\theta, 1-\theta)$ is the loss faced by investors when they
ignore the predictive signals, and $U(\theta)$ is the extent to which
the extreme forecasting model offers better performance than no model at
all \citep{Betz-Oprica-Peltonen-Sarlin-2014-JBF}.  Extreme forecasts are
useful when $U(\theta) > 0$, which means that losses using the forecasts
are lower than when the forecasts are ignored. The usefulness definition
here ignores any influence from the data imbalance, i.e., that
non-extreme events occur much more frequently than extreme events
\citep{Sarlin-2013-EL, Betz-Oprica-Peltonen-Sarlin-2014-JBF}.

Given hazard probability $W(\Delta t |t)$, we need a hazard threshold
$w_t$ that maximizes usefulness $U(\theta)$
\citep{Duca-Peltonen-2013-JBF,
  Babecky-Havranek-Mateju-Rusnak-Smidkova-Vasicek-2014-JFS,
  Betz-Oprica-Peltonen-Sarlin-2014-JBF}. \cite{Christensen-Li-2014-JFS}
optimizes the threshold by minimizing the noise-to-signal ratio
$D/A$. When we optimize the usefulness there is a marginal rate of
substitution between Type 1 and Type 2 errors, but this marginal rate is
not clear in the optimization of the noise-to-signal ratio, and this can
result in an unacceptable level of Type 1 and Type 2 errors
\citep{Alessi-Detken-2011-EJPE, ElShagi-Knedlik-vonSchweinitz-2013-JIMF,
  Babecky-Havranek-Mateju-Rusnak-Smidkova-Vasicek-2014-JFS}.

\subsection{Estimating distributional parameters}
\label{Sec:MM:Fits}

\noindent
By introducing the stretched exponential function of
Eq.~(\ref{Eq:PDF:sExp}) into the probability density function
$\int_0^{+\infty} p(\tau) {\rm{d}}\tau = 1$, we obtain
\begin{equation}\label{Eq:sE:1}
\frac{a}{\mu b} \Gamma (\frac{1}{\mu}) = 1,
\end{equation}
where $\Gamma(x)$ is the Gamma function.
\cite{Podobnik-Horvatic-Petersen-Stanley-2009-PNAS} and
\cite{Bogachev-Bunde-2009-PRE} describe the one-to-one correspondence
between the average recurrence interval $\tau_Q$ and the percentage of
extremes,
\begin{equation}\label{Eq:RI:Mean}
\tau_Q = \frac{1}{\int_m^{+\infty} p_r(r) {\rm{d}} r} = \frac{1}{1 -
  \int_{-\infty}^m p_r(r) {\rm{d}} r} = \frac{1}{1-Q}, 
\end{equation}
where $Q$ is the quantile that is used to define the extreme values. For
this equation to be valid, the extremes must be positive. When extremes
are negative, we convert them into positives by multiplying by
$-1$. \cite{Chicheportiche-Chakraborti-2014-PRE} find that the average
recurrence interval is universal irrespective of the dependence
structure of the underlying process. From the definition of expectation,
the average recurrence interval can also be written $\tau_Q =
\int_0^{+\infty} \tau p(\tau) {\rm{d}}\tau$. For the stretched
exponential distribution, we have
\begin{equation}\label{Eq:sE:2}
\frac{a}{\mu b^2} \Gamma(\frac{2}{\mu}) = \tau_Q.
\end{equation}

By solving Eqs.~(\ref{Eq:sE:1}) and (\ref{Eq:sE:2}) and using $\mu$ and
$\tau_Q$ for the stretched exponential distribution, parameters $a$ and
$b$ are
\begin{equation}\label{Eq:sE:ab:mu}
a = \frac{\mu \Gamma(2/\mu)}{\Gamma(1/\mu)^2 \tau_Q}, ~~ b = \frac{\Gamma(2/\mu)}{\Gamma(1/\mu) \tau_Q}.
\end{equation}
This strategy reduces the number of estimated parameters from three to
one.

The mean of the $q$-exponential distribution is $1/[\lambda(3-2q)]$. For
there to be a mean, $q$ must be less than 3/2.  Then the parameter
$\lambda$ can be found using $q$ and $\tau_Q$,
\begin{equation}\label{Eq:qE:lambda:q}
\lambda = \frac{1}{\tau_Q(3-2q)}.
\end{equation}

The expectation for the Weibull distribution is $\beta
\Gamma(1+1/\alpha)$. Similarly, the $\beta$ can be expressed in terms of
$\alpha$ and $\tau_Q$,
\begin{equation}\label{Eq:qE:beta:alpha}
\beta = \frac{\tau_Q}{\Gamma(1+\frac{1}{\alpha})}.
\end{equation}

We need to estimate only one parameter for the three distributions when
they are used to fit the recurrence intervals. We adopt the maximum
likelihood estimation (MLE) to estimate the distributional
parameters. The logarithmic likelihood functions are the stretched
exponential distribution
\begin{equation}\label{Eq:sE:lnL}
\ln L_{\rm{sE}} = n \ln a - \sum_{i=1}^n (b \tau_i)^{\mu},
\end{equation}
the $q$-exponential distribution
\begin{equation}\label{Eq:qE:lnL}
\ln L_{\rm{qE}} = n \ln[\lambda (2-q)] -\frac{1}{q-1} \sum_{i=1}^{n} \ln
    [1+(q-1)\lambda \tau_i], 
\end{equation}
and the Weibull distribution
\begin{equation}\label{Eq:W:lnL}
\ln L_{\rm{W}} = n \ln \frac{\alpha}{\beta} + \sum_{i=1}^n
\left[(\alpha-1) \ln \frac{\tau_i}{\beta} - \left(\frac{\tau_i}{\beta}
  \right)^\alpha\right], 
\end{equation}
in which $n$ is the number of recurrence intervals.

Taking the stretched exponential distribution as an example, the
logarithmic likelihood function $\ln L_{\rm{sE}}$ is a one-variable
function of $\mu$. Although usually we can solve the equation by taking
the first order derivative of $\ln L_{\rm{sE}}$ with respect to $\mu$ to
find the solution that maximizes the likelihood, here the analytical
expression of the derivative of $\ln L_{\rm{sE}}$ with respect to $\mu$
is more difficult to obtain. We thus discretize $\mu$ in the $(0, 1)$
range with a step of $10^{-6}$ and calculate the logarithmic likelihood
function for each discrete value of $\mu$. The $\mu$ associated with the
maximum value of $\ln L_{\rm{sE}}$ is the maximum likelihood
estimation. In the same way we estimate the parameters $q \in (1, 1.5)$
and $\alpha \in (0, 1)$ for the $q$-exponential and Weibull
distributions, respectively.

\section{Recurrence interval analysis}

\noindent
In order to test the validity of our extreme-return-prediction model, we
use the data before each turbulent period to calibrate the model and
each turbulent period that follows for out-of-sample forecasting. We
obtain six in-sample calibrating periods: 1885--1928, 1885--1972,
1885--1986, 1885--1999, 1885--2006, and 1885--2010 Their out-of-sample
predicting periods are 1929--1932, 1973--1975, 1987--1989, 2000--2003,
2007--2009, and 2011--2015, respectively. In each in-sample calibrating
period, we identify the extreme value threshold $x_t$ and extract the
extreme values associated with $x_t$.  We also locate the extreme values
based on the quantile thresholds of 95\%, 97.5\%, and 99\%. For each
group of extreme values, we estimate the waiting time between
consecutive extreme values, i.e., the recurrence interval. We take
positive, negative, and absolute returns into account in our analysis
because they may have connections with specific trading strategies. The
investors holding long positions in the market are more sensitive to
extreme negative returns and those holding short positions less
sensitive.

Table \ref{Tb:RI:Statistics} lists the recurrence intervals for
different thresholds from different calibrating periods. Note that
unlike the observations from the quantile threshold, the observations
from the extreme value threshold do not increase monotonically as the
calibration periods are expanded. The sharp decease in the number of
recurrence intervals, for example from Panel B to Panel C for positive
returns and from Panel C to Panel D for absolute returns, indicates that
there is a dramatic increase in the extreme value threshold, suggesting
that the market after 1973 became more volatile [see
  Fig.~\ref{Fig:Index:Return}(b)]. The mean values of the recurrence
interval are strongly influenced by the extreme values, as indicated by
the large gap between the means and medians. Because the skewness is
positive and the kurtosis is much greater than 3, the recurrence
intervals also exhibit a right-skewed and fat-tailed distribution. This
affirms the finding that the recurrence intervals obey a stretched
exponential \citep{Xie-Jiang-Zhou-2014-EM, Suo-Wang-Li-2015-EM,
  Jiang-Canabarro-Podobnik-Stanley-Zhou-2016-QF} or a $q$-exponential
distribution \citep{Ludescher-Tsallis-Bunde-2011-EPL,
  Ludescher-Bunde-2014-PRE, Chicheportiche-Chakraborti-2014-PRE}.

We see a significantly positive autocorrelation of lag 1 in the 95\%
column in Panel A and significant Ljung-Box Q statistics at the 0.01
level for the three types of returns. In addition the autocorrelation of
lag 5 is also positive when returns are positive. This indicates that
there are autocorrelations at short and long lags in the recurrence
intervals at the 95\% quantile threshold. In the 99\% column the
autocorrelations are close to 0 and the Ljung-Box Q statistics are
insignificant for positive, negative, and absolute returns, suggesting
that there are no correlations in the recurrence intervals. The results
in both columns also show that the autocorrelation of recurrence
intervals gradually decreases to insignificance when the quantile
threshold increases from 95\% to 99\%, which is also seen in the columns of $x_t$
and 97.5\% quantile. In Panels B through F, the
autocorrelation coefficients at Lags 1 and 5 are all positive and
statistically significant, indicating the presence of strong
autocorrelations in the recurrence intervals. In addition, the Ljung-Box
Q statistics of lag 30 are statistically significant at the 1\% level,
implying that significant autocorrelations are also prevalent when
recurrence interval lags are longer. These results are supported by the
long-memory behavior results of a detrended fluctuation analysis (DFA)
of the recurrence intervals \citep{Ren-Zhou-2010-PRE,
  Xie-Jiang-Zhou-2014-EM, Suo-Wang-Li-2015-EM}.

\begin{table*}[tb]
\setlength\tabcolsep{1.6pt}
\footnotesize
\centering
 \caption{\label{Tb:RI:Statistics} Descriptive statistics of the
   recurrence intervals. This table reports number of observations
   (obsv), mean, median, standard deviation (stdev), skewness (skew),
   kurtosis (kurt), autocorrelation (rho), Ljung-Box Q test statistic
   (LBQ) of the return intervals between the extreme events, defined by
   the extreme value thresholds $x_t$ and quantile thresholds (95\%, 97.5\%, and 99\%). The
   autocorrelation coefficients are calculated at lag 1 and 5. The
   Ljung-Box Q tests are conducted at lag 30. Panels A--F present the
   results from different in-sample calibrating periods.}
 \medskip
 \centering
\begin{tabular}{lcr@{.}lr@{.}lr@{.}lr@{.}lcr@{.}lr@{.}lr@{.}lr@{.}lcr@{.}lr@{.}lr@{.}lr@{.}l}
\toprule
&& \multicolumn{8}{c}{Negative return} && \multicolumn{8}{c}{Positive return} && \multicolumn{8}{c}{Absolute return} \\  %
 \cline{3-10} \cline{12-19}  \cline{21-28}
   &&  \multicolumn{2}{c}{$x_t$} & \multicolumn{2}{c}{95\%} & \multicolumn{2}{c}{97.5\%}  & \multicolumn{2}{c}{99\%}  &&  \multicolumn{2}{c}{$x_t$}  &  \multicolumn{2}{c}{95\%} & \multicolumn{2}{c}{97.5\%}  & \multicolumn{2}{c}{99\%} &&  \multicolumn{2}{c}{$x_t$} &  \multicolumn{2}{c}{95\%} & \multicolumn{2}{c}{97.5\%}  & \multicolumn{2}{c}{99\%} \\
\midrule

\multicolumn{28}{l}{Panel A: in-sample calibrating period (1885 -- 1928)} \\
 obsv && 278&0 & 652&0 & 326&0 & 130&0 && 477&0 & 652&0 & 326&0 & 130&0 && 145&0 & 652&0 & 326&0 & 130&0 \\
 mean && 46&950 & 20&0 & 40&0 & 100&0 && 27&363 & 20&0 & 40&0 & 100&0 && 90&014 & 20&0 & 40&0 & 100&0 \\
 median && 12&0 & 6&500 & 11&0 & 21&0 && 12&0 & 9&0 & 16&0 & 23&500 && 12&0 & 5&0 & 6&500 & 12&0 \\
 stdev && 91&524 & 38&122 & 78&407 & 173&978 && 44&790 & 32&100 & 78&043 & 184&695 && 182&677 & 46&844 & 94&819 & 201&517 \\
 skew && 4&713 & 6&703 & 5&131 & 3&859 && 3&782 & 4&286 & 6&067 & 3&677 && 3&704 & 9&852 & 6&000 & 3&402 \\
 kurt && 33&706 & 80&856 & 43&300 & 24&150 && 20&607 & 27&888 & 55&805 & 20&586 && 20&540 & 157&674 & 57&471 & 16&804 \\
 rho(1) && 0&120$^{**}$ & 0&121$^{***}$ & 0&140$^{**}$ & 0&070 && 0&257$^{***}$ & 0&218$^{***}$ & 0&081 & $-$0&028 && 0&105 & 0&143$^{***}$ & 0&090 & 0&082 \\
 rho(5) && $-$0&064 & 0&062 & $-$0&053 & $-$0&054 && 0&083$^{*}$ & 0&112$^{***}$ & 0&142$^{**}$ & $-$0&086 && $-$0&040 & 0&005 & 0&041 & $-$0&058 \\
 Q(30) && 48&876$^{**}$ & 66&026$^{***}$ & 36&545 & 18&890 && 131&947$^{***}$ & 143&181$^{***}$ & 41&158$^{*}$ & 14&910 && 18&024 & 53&899$^{***}$ & 30&233 & 10&549 \\
\hline 
\multicolumn{28}{l}{Panel B: in-sample calibrating period (1885 -- 1972)} \\
 obsv && 2032&0 & 1254&0 & 627&0 & 250&0 && 1171&0 & 1254&0 & 627&0 & 250&0 && 2546&0 & 1254&0 & 627&0 & 250&0 \\
 mean && 12&349 & 20&0 & 40&0 & 100&0 && 21&430 & 20&0 & 40&0 & 100&0 && 9&856 & 20&0 & 40&0 & 100&0 \\
 median && 4&0 & 5&0 & 6&0 & 8&0 && 7&0 & 7&0 & 8&0 & 12&0 && 3&0 & 3&0 & 4&0 & 6&0 \\
 stdev && 22&546 & 43&451 & 110&804 & 251&118 && 48&351 & 43&793 & 105&520 & 281&113 && 23&928 & 59&488 & 138&074 & 276&742 \\
 skew && 5&010 & 5&606 & 7&684 & 4&409 && 8&208 & 7&289 & 6&213 & 6&072 && 7&690 & 7&306 & 7&189 & 4&007 \\
 kurt && 41&635 & 47&837 & 84&779 & 26&219 && 110&246 & 83&273 & 50&569 & 50&613 && 89&933 & 73&524 & 65&717 & 19&536 \\
 rho(1) && 0&193$^{***}$ & 0&295$^{***}$ & 0&140$^{***}$ & 0&532$^{***}$ && 0&310$^{***}$ & 0&303$^{***}$ & 0&193$^{***}$ & 0&215$^{***}$ && 0&219$^{***}$ & 0&194$^{***}$ & 0&075$^{*}$ & 0&311$^{***}$ \\
 rho(5) && 0&112$^{***}$ & 0&098$^{***}$ & 0&245$^{***}$ & 0&090 && 0&117$^{***}$ & 0&106$^{***}$ & 0&131$^{***}$ & 0&309$^{***}$ && 0&158$^{***}$ & 0&085$^{***}$ & 0&110$^{***}$ & 0&407$^{***}$ \\
 Q(30) && 843&689$^{***}$ & 755&150$^{***}$ & 206&777$^{***}$ & 111&717$^{***}$ && 753&825$^{***}$ & 733&710$^{***}$ & 432&781$^{***}$ & 82&437$^{***}$ && 1153&606$^{***}$ & 585&612$^{***}$ & 225&441$^{***}$ & 110&987$^{***}$ \\
\hline 
\multicolumn{28}{l}{Panel C: in-sample calibrating period (1885 -- 1986)} \\
 obsv && 716&0 & 1431&0 & 715&0 & 286&0 && 1360&0 & 1431&0 & 715&0 & 286&0 && 2822&0 & 1431&0 & 715&0 & 286&0 \\
 mean && 39&989 & 20&0 & 40&0 & 100&0 && 21&053 & 20&0 & 40&0 & 100&0 && 10&146 & 20&0 & 40&0 & 100&0 \\
 median && 7&0 & 5&0 & 7&0 & 8&0 && 8&0 & 7&0 & 9&0 & 12&0 && 3&0 & 4&0 & 4&0 & 6&0 \\
 stdev && 99&845 & 43&138 & 99&905 & 253&139 && 46&590 & 44&787 & 102&571 & 272&215 && 24&320 & 58&219 & 128&684 & 270&823 \\
 skew && 5&139 & 5&507 & 5&135 & 4&222 && 8&144 & 8&664 & 6&099 & 5&977 && 7&815 & 7&182 & 6&986 & 3&832 \\
 kurt && 34&807 & 45&983 & 34&762 & 24&000 && 111&272 & 124&207 & 50&108 & 50&643 && 93&261 & 72&012 & 66&050 & 18&315 \\
 rho(1) && 0&242$^{***}$ & 0&284$^{***}$ & 0&242$^{***}$ & 0&565$^{***}$ && 0&296$^{***}$ & 0&259$^{***}$ & 0&196$^{***}$ & 0&221$^{***}$ && 0&233$^{***}$ & 0&205$^{***}$ & 0&098$^{***}$ & 0&213$^{***}$ \\
 rho(5) && 0&176$^{***}$ & 0&117$^{***}$ & 0&174$^{***}$ & 0&307$^{***}$ && 0&114$^{***}$ & 0&130$^{***}$ & 0&136$^{***}$ & 0&271$^{***}$ && 0&148$^{***}$ & 0&075$^{***}$ & 0&080$^{**}$ & 0&237$^{***}$ \\
 Q(30) && 276&588$^{***}$ & 667&824$^{***}$ & 277&421$^{***}$ & 249&183$^{***}$ && 791&779$^{***}$ & 736&976$^{***}$ & 450&736$^{***}$ & 101&007$^{***}$ && 1277&446$^{***}$ & 597&888$^{***}$ & 238&276$^{***}$ & 182&886$^{***}$ \\
\hline 
\multicolumn{28}{l}{Panel D: in-sample calibrating period (1885 -- 1999)} \\
 obsv && 782&0 & 1595&0 & 797&0 & 319&0 && 1448&0 & 1595&0 & 797&0 & 319&0 && 1421&0 & 1595&0 & 797&0 & 319&0 \\
 mean && 40&816 & 20&0 & 40&0 & 100&0 && 22&043 & 20&0 & 40&0 & 100&0 && 22&462 & 20&0 & 40&0 & 100&0 \\
 median && 8&0 & 6&0 & 7&0 & 8&0 && 8&0 & 8&0 & 9&0 & 13&0 && 4&0 & 4&0 & 4&0 & 5&0 \\
 stdev && 97&920 & 41&851 & 96&353 & 248&548 && 47&351 & 43&650 & 99&684 & 270&301 && 62&523 & 56&713 & 127&547 & 270&274 \\
 skew && 5&055 & 5&511 & 5&155 & 4&210 && 7&660 & 8&550 & 6&076 & 5&746 && 6&589 & 7&148 & 7&004 & 3&903 \\
 kurt && 34&510 & 46&938 & 35&971 & 23&986 && 99&818 & 124&073 & 50&734 & 46&911 && 60&862 & 72&525 & 65&444 & 18&877 \\
 rho(1) && 0&256$^{***}$ & 0&278$^{***}$ & 0&252$^{***}$ & 0&548$^{***}$ && 0&327$^{***}$ & 0&259$^{***}$ & 0&228$^{***}$ & 0&199$^{***}$ && 0&203$^{***}$ & 0&214$^{***}$ & 0&089$^{**}$ & 0&206$^{***}$ \\
 rho(5) && 0&176$^{***}$ & 0&131$^{***}$ & 0&189$^{***}$ & 0&273$^{***}$ && 0&142$^{***}$ & 0&139$^{***}$ & 0&131$^{***}$ & 0&326$^{***}$ && 0&081$^{***}$ & 0&082$^{***}$ & 0&148$^{***}$ & 0&203$^{***}$ \\
 Q(30) && 277&454$^{***}$ & 675&904$^{***}$ & 303&462$^{***}$ & 254&647$^{***}$ && 849&279$^{***}$ & 790&724$^{***}$ & 467&774$^{***}$ & 102&416$^{***}$ && 677&953$^{***}$ & 657&786$^{***}$ & 195&112$^{***}$ & 154&916$^{***}$ \\
\hline 
\multicolumn{28}{l}{Panel E: in-sample calibrating period (1885 -- 2006)} \\
 obsv && 834&0 & 1683&0 & 841&0 & 336&0 && 1539&0 & 1683&0 & 841&0 & 336&0 && 1518&0 & 1683&0 & 841&0 & 336&0 \\
 mean && 40&380 & 20&0 & 40&0 & 100&0 && 21&882 & 20&0 & 40&0 & 100&0 && 22&185 & 20&0 & 40&0 & 100&0 \\
 median && 8&0 & 6&0 & 8&0 & 8&0 && 8&0 & 8&0 & 9&0 & 13&0 && 4&0 & 4&0 & 4&0 & 6&0 \\
 stdev && 95&125 & 41&361 & 94&757 & 241&773 && 46&746 & 43&235 & 102&416 & 265&009 && 60&659 & 56&179 & 124&604 & 264&232 \\
 skew && 5&213 & 5&472 & 5&240 & 4&322 && 7&580 & 8&417 & 6&467 & 5&925 && 6&795 & 7&162 & 7&148 & 3&985 \\
 kurt && 36&645 & 46&813 & 36&980 & 25&403 && 99&236 & 122&461 & 56&734 & 50&015 && 64&680 & 72&654 & 68&333 & 19&714 \\
 rho(1) && 0&258$^{***}$ & 0&295$^{***}$ & 0&254$^{***}$ & 0&536$^{***}$ && 0&337$^{***}$ & 0&286$^{***}$ & 0&282$^{***}$ & 0&199$^{***}$ && 0&206$^{***}$ & 0&221$^{***}$ & 0&102$^{***}$ & 0&204$^{***}$ \\
 rho(5) && 0&178$^{***}$ & 0&105$^{***}$ & 0&193$^{***}$ & 0&278$^{***}$ && 0&151$^{***}$ & 0&141$^{***}$ & 0&081$^{**}$ & 0&251$^{***}$ && 0&083$^{***}$ & 0&091$^{***}$ & 0&058$^{*}$ & 0&204$^{***}$ \\
 Q(30) && 304&082$^{***}$ & 873&194$^{***}$ & 304&969$^{***}$ & 260&485$^{***}$ && 980&303$^{***}$ & 875&671$^{***}$ & 491&718$^{***}$ & 106&786$^{***}$ && 743&033$^{***}$ & 673&330$^{***}$ & 200&311$^{***}$ & 158&391$^{***}$ \\
\hline 
\multicolumn{28}{l}{Panel F:in-sample calibrating period (1885 -- 2010)} \\
 obsv && 858&0 & 1734&0 & 867&0 & 346&0 && 924&0 & 1734&0 & 867&0 & 346&0 && 1540&0 & 1734&0 & 867&0 & 346&0 \\
 mean && 40&425 & 20&0 & 40&0 & 100&0 && 37&538 & 20&0 & 40&0 & 100&0 && 22&523 & 20&0 & 40&0 & 100&0 \\
 median && 7&0 & 6&0 & 7&0 & 8&0 && 9&0 & 8&0 & 9&0 & 12&0 && 4&0 & 4&0 & 4&0 & 5&0 \\
 stdev && 111&346 & 42&218 & 109&226 & 259&435 && 97&671 & 43&424 & 104&937 & 269&598 && 68&579 & 59&451 & 132&133 & 275&164 \\
 skew && 7&095 & 5&598 & 7&081 & 4&464 && 6&397 & 8&211 & 6&287 & 5&680 && 8&097 & 7&761 & 6&888 & 3&898 \\
 kurt && 71&630 & 49&644 & 72&329 & 26&510 && 56&115 & 117&442 & 52&915 & 46&183 && 91&999 & 83&058 & 61&436 & 18&701 \\
 rho(1) && 0&146$^{***}$ & 0&296$^{***}$ & 0&161$^{***}$ & 0&495$^{***}$ && 0&308$^{***}$ & 0&297$^{***}$ & 0&302$^{***}$ & 0&184$^{***}$ && 0&218$^{***}$ & 0&201$^{***}$ & 0&112$^{***}$ & 0&165$^{***}$ \\
 rho(5) && 0&174$^{***}$ & 0&088$^{***}$ & 0&132$^{***}$ & 0&234$^{***}$ && 0&084$^{**}$ & 0&138$^{***}$ & 0&191$^{***}$ & 0&218$^{***}$ && 0&113$^{***}$ & 0&099$^{***}$ & 0&057$^{*}$ & 0&249$^{***}$ \\
 Q(30) && 227&668$^{***}$ & 877&559$^{***}$ & 215&488$^{***}$ & 223&093$^{***}$ && 552&569$^{***}$ & 994&993$^{***}$ & 465&599$^{***}$ & 99&923$^{***}$ && 565&484$^{***}$ & 613&619$^{***}$ & 218&575$^{***}$ & 146&530$^{***}$ \\

\bottomrule
\end{tabular}
\end{table*}

\begin{figure}[htbp]
 \centering
  \includegraphics[width=8cm]{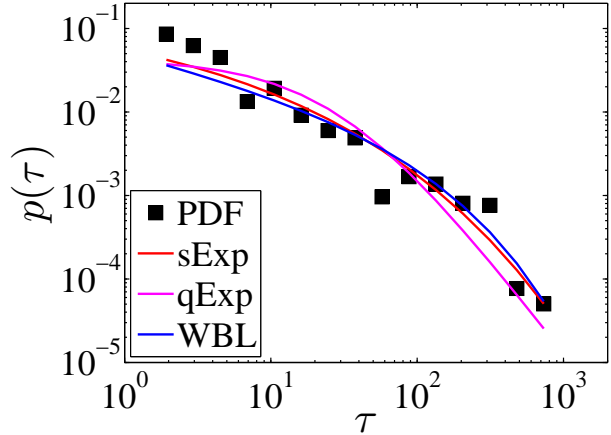}
 \caption{\label{Fig:Distribution:RI:NR:QT99} (color
   online). Distribution of the recurrence intervals. This figure
   presents the empirical distributions of the recurrence intervals
   between negative extreme returns in the quantile of 99\% and the best
   fits to the three distributions, stretched exponential distribution,
   $q$-exponential distribution, and Weibull distribution. The analysis
   is performed in the period from 1885 to 1928.}
\end{figure}

The recurrence intervals are fitted by the stretched exponential distribution,
$q$-exponential distribution, and Weibull distribution in each in-sample
calibrating period. Figure~\ref{Fig:Distribution:RI:NR:QT99} shows the
probability distribution of the recurrence intervals between the
negative extreme events in the 99\% quantile during the 1928 to 1985
in-sample calibrating period. The best fits to the three fitting
distributions are also plotted as solid curves for comparison. Note that
the stretched exponential distribution gives the best fit. Note also that
the stretch exponential fits are most likely, which also agrees with the
distribution of recurrence intervals between the negative and absolute
extreme returns in the Chinese markets and the US markets
\citep{Wang-Yamasaki-Havlin-Stanley-2009-PRE, Ren-Zhou-2010-NJP,
  Xie-Jiang-Zhou-2014-EM, Suo-Wang-Li-2015-EM}. Because all the
distribution curves are very similar, we do not show the recurrence
intervals in other calibrating periods.

Table~\ref{Tb:RI:Fit:Par} shows the estimated parameters of three
fitting distributions of the recurrence intervals obtained from
different return types and different thresholds. To assess the validity
of the distributional fits, the logarithmic likelihoods are also
listed. The likelihoods of the distribution with the maximum value are
show in bold to indicate that the corresponding distribution gives the
best fit. Panel A shows that the best recurrence interval fits
transition from the $q$-exponential distribution to the stretched
exponential distribution as the threshold is increased. This
distribution transition behavior is also seen in the recurrence
intervals in the minute volatilities in the Chinese stock markets
\citep{Jiang-Canabarro-Podobnik-Stanley-Zhou-2016-QF}. Note that in
Panels B--F the maximum likelihood comes from the $q$-exponential
distribution for all the recurrence interval
fits. \cite{Ludescher-Tsallis-Bunde-2011-EPL} and
\cite{Ludescher-Bunde-2014-PRE} also found that the recurrence intervals
between extreme loses are captured by the $q$-exponential
distribution. The possible explanation for the lack of maximum
likelihoods for the stretched exponential distribution in Panels B--F is
that the 99\% quantile threshold is not sufficiently large for a
distribution transition from a $q$-exponential to a stretched
exponential to occur.

Table~\ref{Tb:RI:Fit:Par} shows a monotonic trend between the estimated
parameters and the quantile thresholds for each type of returns, such
that $\mu$ and $\alpha$ decrease and $q$ increases as the quantile
threshold increases. Our results support the dependence of the
recurrence interval distribution on the quantile threshold
\citep{Xie-Jiang-Zhou-2014-EM, Chicheportiche-Chakraborti-2014-PRE,
  Suo-Wang-Li-2015-EM,
  Jiang-Canabarro-Podobnik-Stanley-Zhou-2016-QF}. For the same quantile
threshold and the same type of return, we also find that the estimated
distributional parameters are close to each other in Panels B--F, which
indicates that the recurrence interval distribution depends solely on
the quantile \citep{Ludescher-Tsallis-Bunde-2011-EPL,
  Ludescher-Bunde-2014-PRE, Jiang-Canabarro-Podobnik-Stanley-Zhou-2016-QF}.

\begin{table*}[tb]
\setlength\tabcolsep{2pt}
\footnotesize
\centering
 \caption{\label{Tb:RI:Fit:Par} The estimated parameters and the maximum
   logarithmic likelihoods of the fits to the stretched exponential
   distribution, $q$-exponential distribution, and Weibull distribution
   for recurrence intervals. The likelihoods with the maximum value are
   highlighted in bold. The recurrence intervals are determined
   according to the extreme value thresholds $x_t$ and quantile
   thresholds (95\%, 97.5\%, and 99\%) in the negative, positive, and
   absolute returns. Panels A--F present the results from different
   in-sample calibrating periods.}
 \medskip
 \centering
\begin{tabular}{lcr@{.}lr@{.}lr@{.}lr@{.}lcr@{.}lr@{.}lr@{.}lr@{.}lcr@{.}lr@{.}lr@{.}lr@{.}l}
\toprule
&& \multicolumn{8}{c}{Negative return} && \multicolumn{8}{c}{Positive return} && \multicolumn{8}{c}{Absolute return} \\  %
 \cline{3-10} \cline{12-19}  \cline{21-28}
   &&  \multicolumn{2}{c}{$x_t$} & \multicolumn{2}{c}{95\%} & \multicolumn{2}{c}{97.5\%}  & \multicolumn{2}{c}{99\%}  &&  \multicolumn{2}{c}{$x_t$}  &  \multicolumn{2}{c}{95\%} & \multicolumn{2}{c}{97.5\%}  & \multicolumn{2}{c}{99\%} &&  \multicolumn{2}{c}{$x_t$} &  \multicolumn{2}{c}{95\%} & \multicolumn{2}{c}{97.5\%}  & \multicolumn{2}{c}{99\%} \\
\midrule
\multicolumn{28}{l}{Panel A: in-sample calibrating period (1885 -- 1928)} \\
 $\mu$ && 0&350 & 0&461 & 0&360 & 0&280 && 0&513 & 0&573 & 0&451 & 0&307 && 0&227 & 0&383 & 0&284 & 0&215 \\
 $\ln L_{\rm{sE}}$ && {\bf $-$1264}&{\bf 269} & $-$2496&914 & {\bf $-$1434}&{\bf 225} & {\bf $-$674}&{\bf 465} && $-$2000&951 & $-$2553&728 & $-$1473&376 & {\bf $-$680}&{\bf 297} && {\bf $-$700}&{\bf 468} & $-$2424&402 & $-$1365&008 & {\bf $-$634}&{\bf 993} \\
 $q$ && 1&408 & 1&357 & 1&405 & 1&435 && 1&316 & 1&286 & 1&345 & 1&423 && 1&461 & 1&399 & 1&443 & 1&465 \\
 $\ln L_{\rm{qE}}$ && $-$1266&953 & {\bf $-$2485}&{\bf 817} & $-$1434&396 & $-$684&927 && {\bf $-$1996}&{\bf 519} & {\bf $-$2544}&{\bf 451} & {\bf $-$1471}&{\bf 303} & $-$685&261 && $-$711&080 & {\bf $-$2411}&{\bf 504} & {\bf $-$1360}&{\bf 696} & $-$647&154 \\
 $\alpha$ && 0&625 & 0&718 & 0&634 & 0&563 && 0&757 & 0&802 & 0&708 & 0&587 && 0&498 & 0&651 & 0&556 & 0&483 \\
 $\ln L_{\rm{W}}$ && $-$1275&718 & $-$2523&094 & $-$1448&428 & $-$678&027 && $-$2014&244 & $-$2571&295 & $-$1483&428 & $-$684&853 && $-$707&857 & $-$2458&188 & $-$1386&102 & $-$641&552 \\
\hline 
\multicolumn{28}{l}{Panel B: in-sample calibrating period (1885 -- 1972)} \\
 $\mu$ && 0&512 & 0&398 & 0&294 & 0&207 && 0&433 & 0&451 & 0&315 & 0&229 && 0&452 & 0&312 & 0&230 & 0&177 \\
 $\ln L_{\rm{sE}}$ && $-$6870&875 & $-$4662&711 & $-$2602&562 & $-$1173&240 && $-$4501&362 & $-$4759&720 & $-$2654&990 & $-$1205&945 && $-$7787&724 & $-$4359&017 & $-$2384&191 & $-$1098&368 \\
 $q$ && 1&336 & 1&396 & 1&442 & 1&473 && 1&373 & 1&363 & 1&429 & 1&464 && 1&373 & 1&438 & 1&467 & 1&483 \\
  $\ln L_{\rm{qE}}$ && $-${\bf 6805} &{\bf 955} & $-${\bf 4610} &{\bf 407} & $-${\bf 2571} &{\bf 429} & $-${\bf 1165} &{\bf 118} && $-${\bf 4463} &{\bf 297} & $-${\bf 4718} &{\bf 507} & $-${\bf 2632} &{\bf 444} & $-${\bf 1200} &{\bf 464} && $-${\bf 7624} &{\bf 907} & $-${\bf 4248} &{\bf 552} & $-${\bf 2312} &{\bf 401} & $-${\bf 1081} &{\bf 346} \\
 $\alpha$ && 0&758 & 0&663 & 0&563 & 0&465 && 0&692 & 0&706 & 0&585 & 0&492 && 0&704 & 0&577 & 0&487 & 0&422 \\
 $\ln L_{\rm{W}}$ && $-$6957&774 & $-$4735&734 & $-$2650&054 & $-$1195&844 && $-$4559&812 & $-$4820&370 & $-$2696&831 & $-$1226&002 && $-$7951&129 & $-$4469&435 & $-$2453&110 & $-$1126&514 \\
\hline 
\multicolumn{28}{l}{Panel C: in-sample calibrating period (1885 -- 1986)} \\
 $\mu$ && 0&302 & 0&409 & 0&302 & 0&208 && 0&449 & 0&463 & 0&325 & 0&229 && 0&464 & 0&322 & 0&237 & 0&179 \\
 $\ln L_{\rm{sE}}$ && $-$3011&043 & $-$5359&481 & $-$3008&233 & $-$1348&487 && $-$5236&811 & $-$5455&453 & $-$3056&744 & $-$1382&459 && $-$8773&031 & $-$5036&183 & $-$2770&080 & $-$1272&260 \\
 $q$ && 1&437 & 1&388 & 1&437 & 1&473 && 1&362 & 1&354 & 1&424 & 1&464 && 1&365 & 1&433 & 1&465 & 1&482 \\
 $\ln L_{\rm{qE}}$ && $-${\bf 2980} &{\bf 534} & $-${\bf 5305} &{\bf 811} & $-${\bf 2978} &{\bf 056} & $-${\bf 1342} &{\bf 692} && $-${\bf 5195} &{\bf 029} & $-${\bf 5410} &{\bf 298} & $-${\bf 3035} &{\bf 525} & $-${\bf 1377} &{\bf 135} && $-${\bf 8608} &{\bf 555} & $-${\bf 4924} &{\bf 461} & $-${\bf 2701} &{\bf 469} & $-${\bf 1258} &{\bf 164} \\
 $\alpha$ && 0&571 & 0&672 & 0&571 & 0&465 && 0&705 & 0&715 & 0&594 & 0&493 && 0&714 & 0&587 & 0&496 & 0&426 \\
 $\ln L_{\rm{W}}$ && $-$3062&519 & $-$5438&722 & $-$3059&523 & $-$1373&600 && $-$5300&912 & $-$5521&812 & $-$3101&494 & $-$1405&021 && $-$8944&445 & $-$5155&682 & $-$2843&627 & $-$1302&926 \\
\hline 
\multicolumn{28}{l}{Panel D: in-sample calibrating period (1885 -- 1999)} \\
 $\mu$ && 0&307 & 0&419 & 0&308 & 0&208 && 0&457 & 0&477 & 0&335 & 0&233 && 0&314 & 0&329 & 0&241 & 0&178 \\
 $\ln L_{\rm{sE}}$ && $-$3325&185 & $-$6004&925 & $-$3375&885 & $-$1506&977 && $-$5657&201 & $-$6104&959 & $-$3430&370 & $-$1552&323 && $-$5155&298 & $-$5655&759 & $-$3115&198 & $-$1416&369 \\
 $q$ && 1&435 & 1&382 & 1&434 & 1&472 && 1&357 & 1&345 & 1&418 & 1&462 && 1&435 & 1&429 & 1&463 & 1&482 \\
 $\ln L_{\rm{qE}}$ && $-${\bf 3298} &{\bf 423} & $-${\bf 5953} &{\bf 468} & $-${\bf 3348} &{\bf 336} & $-${\bf 1501} &{\bf 836} && $-${\bf 5615} &{\bf 392} & $-${\bf 6055} &{\bf 245} & $-${\bf 3408} &{\bf 686} & $-${\bf 1548} &{\bf 247} && $-${\bf 5054} &{\bf 298} & $-${\bf 5540} &{\bf 644} & $-${\bf 3047} &{\bf 366} & $-${\bf 1403} &{\bf 017} \\
 $\alpha$ && 0&577 & 0&681 & 0&578 & 0&467 && 0&711 & 0&726 & 0&604 & 0&498 && 0&581 & 0&595 & 0&502 & 0&424 \\
 $\ln L_{\rm{W}}$ && $-$3378&117 & $-$6087&515 & $-$3429&832 & $-$1534&442 && $-$5722&565 & $-$6175&819 & $-$3477&890 & $-$1576&590 && $-$5271&119 & $-$5783&800 & $-$3193&905 & $-$1450&209 \\
\hline 
\multicolumn{28}{l}{Panel E: in-sample calibrating period (1885 -- 2006)} \\
 $\mu$ && 0&310 & 0&422 & 0&311 & 0&214 && 0&455 & 0&472 & 0&334 & 0&235 && 0&318 & 0&331 & 0&241 & 0&181 \\
 $\ln L_{\rm{sE}}$ && $-$3529&464 & $-$6339&332 & $-$3552&077 & $-$1598&990 && $-$5994&090 & $-$6429&892 & $-$3604&772 & $-$1636&240 && $-$5476&112 & $-$5959&563 & $-$3276&784 & $-$1502&338 \\
 $q$ && 1&434 & 1&381 & 1&433 & 1&470 && 1&359 & 1&350 & 1&419 & 1&461 && 1&434 & 1&429 & 1&463 & 1&481 \\
  $\ln L_{\rm{qE}}$ && $-${\bf 3498} &{\bf 388} & $-${\bf 6285} &{\bf 001} & $-${\bf 3519} &{\bf 699} & $-${\bf 1595} &{\bf 926} && $-${\bf 5947} &{\bf 393} & $-${\bf 6375} &{\bf 880} & $-${\bf 3577} &{\bf 750} & $-${\bf 1632} &{\bf 522} && $-${\bf 5364} &{\bf 948} & $-${\bf 5835} &{\bf 504} & $-${\bf 3203} &{\bf 838} & $-${\bf 1491} &{\bf 093} \\
 $\alpha$ && 0&582 & 0&684 & 0&582 & 0&475 && 0&710 & 0&723 & 0&603 & 0&501 && 0&585 & 0&597 & 0&503 & 0&430 \\
 $\ln L_{\rm{W}}$ && $-$3585&778 & $-$6425&731 & $-$3609&157 & $-$1626&494 && $-$6064&641 & $-$6506&061 & $-$3656&471 & $-$1661&245 && $-$5599&175 & $-$6094&643 & $-$3359&717 & $-$1536&739 \\
\hline 
\multicolumn{28}{l}{Panel F: in-sample calibrating period (1885 -- 2010)} \\
 $\mu$ && 0&305 & 0&421 & 0&308 & 0&208 && 0&338 & 0&469 & 0&333 & 0&232 && 0&312 & 0&330 & 0&237 & 0&177 \\
 $\ln L_{\rm{sE}}$ && $-$3616&004 & $-$6527&378 & $-$3651&908 & $-$1632&635 && $-$3910&386 & $-$6620&877 & $-$3712&984 & $-$1682&142 && $-$5558&880 & $-$6135&844 & $-$3356&883 & $-$1529&646 \\
 $q$ && 1&436 & 1&382 & 1&435 & 1&472 && 1&418 & 1&352 & 1&421 & 1&462 && 1&437 & 1&429 & 1&465 & 1&483 \\
  $\ln L_{\rm{qE}}$ && $-${\bf 3577} &{\bf 687} & $-${\bf 6465} &{\bf 676} & $-${\bf 3613} &{\bf 537} & $-${\bf 1625} &{\bf 505} && $-${\bf 3875} &{\bf 292} & $-${\bf 6562} &{\bf 953} & $-${\bf 3680} &{\bf 043} & $-${\bf 1677} &{\bf 062} && $-${\bf 5441} &{\bf 337} & $-${\bf 6002} &{\bf 342} & $-${\bf 3273} &{\bf 580} & $-${\bf 1512} &{\bf 333} \\
 $\alpha$ && 0&574 & 0&682 & 0&577 & 0&466 && 0&606 & 0&720 & 0&601 & 0&497 && 0&577 & 0&595 & 0&497 & 0&422 \\
 $\ln L_{\rm{W}}$ && $-$3678&097 & $-$6619&216 & $-$3714&149 & $-$1663&022 && $-$3969&370 & $-$6701&078 & $-$3768&877 & $-$1708&772 && $-$5688&562 & $-$6278&192 & $-$3446&235 & $-$1567&267 \\
\bottomrule
\end{tabular}
\end{table*}


When we obtain the distribution parameters, we can then find the
theoretical curve of the hazard function for the $W_{\rm{sE}}$,
$W_{\rm{qW}}$, and $W_{\rm{W}}$ by putting the parameters
into the theoretical formula for hazard probability $W(\Delta t | t)$
given by Eqs.~(\ref{Eq:sExp:Hazard}), (\ref{Eq:qExp:Hazard}), and
(\ref{Eq:WBL:Hazard}) for the stretched exponential distribution,
$q$-exponential distribution, and Weibull distribution, respectively.
On the other hand, using Eq.~(\ref{Eq:Wq}) we can evaluate the empirical
hazard function $W_{\rm{emp}}$,
\begin{equation}\label{Eq:emp:Hazard}
W_{\rm{emp}}(\Delta t |t) = \frac{\#(t < \tau \le t+\Delta t)}{ \#(\tau > t)},
\end{equation}
where the denominator $\#(\tau>t)$ is the number of recurrence intervals
with values greater than $t$, and the numerator $\#( t < \tau \le t +
\Delta t)$ the number of recurrence intervals within the range of $(t,
t+\Delta t]$.

\begin{figure}[t]
 \centering
  \includegraphics[width=8cm]{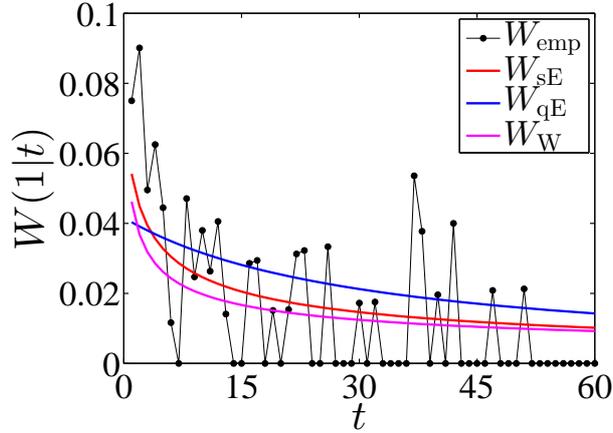}
  \caption{\label{Fig:HP:RI:W} (color online). Plots of hazard
    probability $W(\Delta t|t)$ with $\Delta t = 1$. The hazard events
    correspond to the extreme negative returns obtained from the
    quantile threshold of 99\%. The analysis is performed in the period
    from 1885 to 1928.}
\end{figure}

Figure~\ref{Fig:HP:RI:W} shows a plot of the hazard probability
$W(\Delta t |t)$ as a function of the elapsing time $t$ for the extreme
negative returns obtained from the 99\% quantile threshold when $\Delta
t =1$. It shows the empirical hazard probability estimated from the real
data (filled markers) and the analytical hazard probabilities obtained
from the theoretical equations (solid curves). Note that although all
the theoretical lines do not overlap on the same curve they all decrease
with respect to the elapsing time $t$, as does the empirical hazard
probability. The statistics are poor and the empirical hazard
probability strongly oscillates, but for a given value of $t$ the
analytical hazard probability values are comparable to those of the
empirical hazard probability, suggesting that the analytical hazard
probabilities agree with the empirical hazard probability. These
decreasing patterns in the hazard probability are also seen in energy
futures \citep{Xie-Jiang-Zhou-2014-EM}, spot index and index futures
\citep{Suo-Wang-Li-2015-EM}, and stock returns \citep{Ren-Zhou-2010-NJP,
  Jiang-Canabarro-Podobnik-Stanley-Zhou-2016-QF}, indicating that the
probability of observing a follow-up extreme return decreases as time
$t$ elapses. This reveals the existence of extreme return clustering and
a potential dependent structure in the triggering processes of the
extreme returns, which supports the argument that ``many extreme price
movements are triggered by previous extreme movements'' and that
``larger extremes occur more often after big events or frequent events
than after tranquil periods''
\citep{Gresnigt-Kole-Franses-2015-JBF}. This is caused by the positive
herding behavior of investors and the endogenous growth of instability
in financial markets
\citep{Jiang-Zhou-Sornette-Woodard-Bastiaensen-Cauwels-2010-JEBO,
  Gresnigt-Kole-Franses-2015-JBF}. Because the results are all similar,
we do not show the hazard probabilities $W(\Delta t |t)$ for different
thresholds and other types of return.

\section{Predicting extreme returns}

\noindent
Using hazard probabilities and an optimized hazard threshold, we build a
model to predict the occurrence of positive, negative, and absolute
extreme returns in financial markets within a given time period. The
hazard probabilities are specified by the distribution parameter of the
recurrence intervals between extreme events in the return history. The
indicators of incoming extreme events are generated when the hazard
probability exceeds the optimized hazard threshold, and this maximizes
the usefulness of these extreme forecasts. We perform out-of-sample
tests to evaluate the predictive power of this extreme-return-prediction
model as follows.

\begin{enumerate}

\item We mark extreme events according to a specified extreme value or
  quantile threshold during a given in-sample calibrating period.
  
\item Fitting the recurrence intervals between the marked extreme
  events, we estimate the stretched exponential distribution,
  $q$-exponential distribution, or Weibull distribution parameters.
    
\item Using the estimated distribution parameters in the in-sample
  calibrating period, we determine the hazard probability $W(\Delta t |
  t)$ and find the optimized hazard threshold $w_t$ by maximizing the
  usefulness $U(\theta)$.
  
\item Using the distribution parameters and optimized hazard threshold
  from the in-sample calibrating period, we forecast the indicators of
  incoming extreme events within time period $\Delta t$ and evaluate the
  forecasting signals.

\end{enumerate}

\begin{figure}[htbp]
 \centering
  \includegraphics[width=8cm]{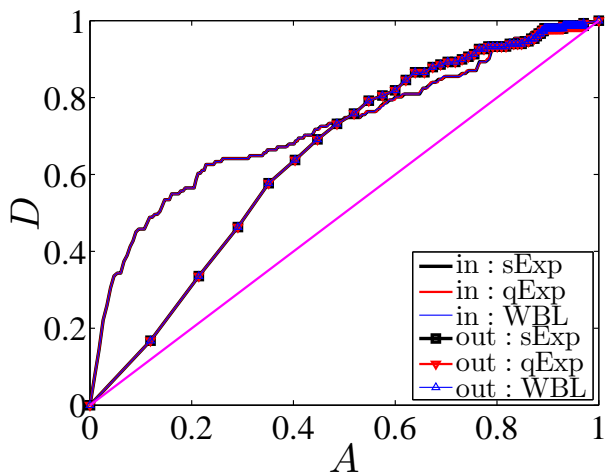}
 \caption{\label{Fig:ROC:inSample} (color online). ROC curves of
   in-sample tests and out-of-sample predictions. The extreme returns
   correspond to the negative returns in the quantile of 99\%. The
   in-sample period covers from 1885 to 1928. The out-of-sample period
   spans from 1929 to 1932.}
\end{figure}

To find the optimized hazard threshold, we vary the hazard threshold in
$[0, 1]$ to obtain all possible pairs of $(A, D)$.  Plotting $A$ with
respect to $D$, we obtain the well-known ``receiver operator
characteristic'' (ROC) curve \citep{Bogachev-Bunde-2009-PRE}. Using the
ROC curve we measure the validity of the predicting power of early
warning models. Figure~\ref{Fig:ROC:inSample} shows the ROC predictive
curves of extreme negative returns for in-sample tests and out-of-sample
tests. The in-sample (out-of-sample) period is from 1885 to 1928 (from
1929 to 1932). The diagonal line is a random guess. Note that the ROC
curves of the three fitting distributions are overlap exactly on the same
curve for in-sample and out-of-sample tests, suggesting that the results
do not depend on the distribution formula used to fit the recurrence
intervals. All ROC curves are above the random guess line, indicating
that both in-sample and out-of-sample tests have a better predictive
power than a random guess. Note also that the out-of-sample curves are
lower than the in-sample curves, which confirms the observation that
out-of-sample predictions are usually worse than in-sample tests
\citep{Lang-Schmidt-2016-JIMF}. Because they all exhibit very similar
patterns, we do not show the ROC curves obtained from different
thresholds and different types of extreme returns.

Because all three fitting distributions give the same ROC curve, we
evaluate only the in-sample and out-of-sample performance of the extreme
return prediction model for the $q$-exponential distribution. We find
the optimized hazard threshold, which maximizes the usefulness in the
in-sample calibrating period, and estimate such performance measurements
as the rate of correct predictions, the false alarm rate, the
usefulness, and the KSS score during in-sample and out-of-sample
periods. The results are shown in Table \ref{Tb:PE:Performance}.

First, we observe that all usefulness $U$ values are positive except in
the positive returns in the 97.5\% quantile in Panel A and in the 95\%
quantile in Panel B, indicating that when missing-event and false-alarm
errors are weighted equally our model provides more accurate results
than the benchmark of ignoring the forecasting signals. Second,
excluding the above two exceptions all KSS scores are greater than 0,
which corresponds to random guessing, indicating that the rate of
correct predictions exceeds that of false alarms. Third, note that in
most of the results, the $U$ and KSS scores of in-sample performance are
larger than those of out-of-sample performance, which is in consistent
with the observation that out-of-sample predictions are inferior to the
in-sample tests. We do find one exception in Panel E and nine exceptions
in Panel F in which the out-of-sample predictions surpass the in-sample
tests, and this indicates the predictive power of the testing model. The
results also imply that the more data available for the in-sample tests,
the better the performance of out-of-sample predictions, and this is
further supported by the predictions during two recent turbulent
periods, which were better than predictions during other
periods. Fourth, note that the predictions of the extreme returns in the
99\% quantile produce a lower false alarm rate and a higher correct
prediction rate than those in the 95\% and 97.5\% quantiles, and this
produces high usefulness and KSS scores. The results imply that the
extreme events with a high quantile can be predicted more accurately.
Table~\ref{Tb:RI:Statistics} shows the statistics of Ljung-Box Q tests
that exhibit a decreasing pattern as quantile thresholds increase in all
panels, indicating that increasing the quantile threshold could decrease
the memory strength in the extremes.  We neglect the potential
dependence structure in the extreme series in our model because the
larger the quantile threshold, the weaker the memory in extremes and the
better the forecasting performance. Compared to the model based on the
Hawkes processes \citep{Gresnigt-Kole-Franses-2015-JBF} our model has
the advantage of fewer model parameters, easier estimating methods, and
a faster prediction implementation.

\begin{table*}[tb]
\setlength\tabcolsep{8.4pt}
\footnotesize
\centering
 \caption{\label{Tb:PE:Performance} In-sample and out-of-sample
   performance of the extreme return predicting model. False alarm
   rates, correct predicting rate, usefulness, and KSS score is listed
   for in-sample tests and out-of-sample predictions. The predictions
   that out-of-sample performances are better than in-sample
   performances are highlighted in bold. The recurrence intervals are
   determined according to the extreme value thresholds $x_t$ and
   quantile thresholds (95\%, 97.5\%, and 99\%) in the negative,
   positive, and absolute returns. Panels A--F present the results
   from different in-sample calibrating periods and out-of-sample
   predicting periods.}
 \medskip
 \centering
\begin{tabular}{ccr@{.}lr@{.}lr@{.}lr@{.}lcr@{.}lr@{.}lr@{.}lr@{.}lcr@{.}lr@{.}lr@{.}lr@{.}l}
\toprule
&& \multicolumn{8}{c}{Negative return} && \multicolumn{8}{c}{Positive return} && \multicolumn{8}{c}{Absolute return} \\  %
 \cline{3-10} \cline{12-19}  \cline{21-28}
  &&  \multicolumn{2}{c}{EVT} & \multicolumn{2}{c}{95\%} & \multicolumn{2}{c}{97.5\%}  & \multicolumn{2}{c}{99\%}  &&  \multicolumn{2}{c}{EVT}  &  \multicolumn{2}{c}{95\%} & \multicolumn{2}{c}{97.5\%}  & \multicolumn{2}{c}{99\%} &&  \multicolumn{2}{c}{EVT} &  \multicolumn{2}{c}{95\%} & \multicolumn{2}{c}{97.5\%}  & \multicolumn{2}{c}{99\%} \\
\midrule
\multicolumn{28}{l}{Panel A: in-sample calibrating period (1885 -- 1928) / out-of-sample predicting period (1929 -- 1932) } \\
 in: $A$ && 0&256 & 0&319 & 0&286 & 0&228 && 0&317 & 0&382 & 0&494 & 0&263 && 0&159 & 0&207 & 0&242 & 0&142 \\
 out: $A$ && 0&807 & 0&814 & 0&834 & 0&895 && 0&821 & 0&862 & 0&967 & 0&773 && 0&719 & 0&699 & 0&805 & 0&721 \\
 in: $D$ && 0&616 & 0&623 & 0&657 & 0&626 && 0&543 & 0&602 & 0&755 & 0&656 && 0&637 & 0&582 & 0&713 & 0&626 \\
 out: $D$ && 0&953 & 0&949 & 0&967 & 0&980 && 0&891 & 0&898 & 0&963 & 0&963 && 0&960 & 0&935 & 0&975 & 0&958 \\
 in: $U$ && 0&180 & 0&152 & 0&186 & 0&199 && 0&113 & 0&110 & 0&131 & 0&197 && 0&239 & 0&187 & 0&235 & 0&242 \\
 out: $U$ && 0&073 & 0&068 & 0&066 & 0&042 && 0&035 & 0&018 & $-$0&002 & 0&095 && 0&121 & 0&118 & 0&085 & 0&119 \\
 in: KSS && 0&361 & 0&304 & 0&372 & 0&398 && 0&226 & 0&220 & 0&261 & 0&393 && 0&478 & 0&375 & 0&470 & 0&484 \\
 out: KSS && 0&146 & 0&135 & 0&133 & 0&084 && 0&069 & 0&036 & $-$0&005 & 0&190 && 0&241 & 0&236 & 0&170 & 0&238 \\
\hline 
\multicolumn{28}{l}{Panel B: in-sample calibrating period (1885 -- 1972) / out-of-sample predicting period (1973 -- 1975) } \\
 in: $A$ && 0&413 & 0&332 & 0&263 & 0&120 && 0&316 & 0&337 & 0&257 & 0&191 && 0&303 & 0&170 & 0&159 & 0&075 \\
 out: $A$ && 0&760 & 0&634 & 0&377 & 0&111 && 0&677 & 0&676 & 0&608 & 0&361 && 0&641 & 0&360 & 0&245 & 0&042 \\
 in: $D$ && 0&727 & 0&729 & 0&760 & 0&725 && 0&657 & 0&664 & 0&702 & 0&757 && 0&723 & 0&700 & 0&799 & 0&737 \\
 out: $D$ && 0&824 & 0&750 & 0&769 & 0&250 && 0&776 & 0&782 & 0&714 & 0&692 && 0&780 & 0&671 & 0&789 & 0&667 \\
 in: $U$ && 0&157 & 0&199 & 0&248 & 0&302 && 0&170 & 0&163 & 0&222 & 0&283 && 0&210 & 0&265 & 0&320 & 0&331 \\
 out: $U$ && 0&032 & 0&058 & 0&196 & 0&069 && 0&049 & 0&053 & 0&053 & 0&166 && 0&069 & 0&155 & 0&272 & 0&312 \\
 in: KSS && 0&314 & 0&397 & 0&497 & 0&605 && 0&341 & 0&327 & 0&445 & 0&566 && 0&420 & 0&531 & 0&640 & 0&663 \\
 out: KSS && 0&064 & 0&116 & 0&392 & 0&139 && 0&099 & 0&106 & 0&107 & 0&331 && 0&139 & 0&311 & 0&545 & 0&624 \\
\hline 
\multicolumn{28}{l}{Panel C: in-sample calibrating period (1885 -- 1986) / out-of-sample predicting period (1987 -- 1989) } \\
 in: $A$ && 0&284 & 0&342 & 0&284 & 0&193 && 0&405 & 0&362 & 0&309 & 0&191 && 0&309 & 0&177 & 0&167 & 0&099 \\
 out: $A$ && 0&370 & 0&512 & 0&370 & 0&291 && 0&721 & 0&688 & 0&676 & 0&394 && 0&546 & 0&336 & 0&229 & 0&205 \\
 in: $D$ && 0&763 & 0&717 & 0&763 & 0&780 && 0&726 & 0&675 & 0&732 & 0&753 && 0&703 & 0&680 & 0&785 & 0&749 \\
 out: $D$ && 0&786 & 0&667 & 0&786 & 0&750 && 0&763 & 0&683 & 0&789 & 0&733 && 0&645 & 0&662 & 0&828 & 0&733 \\
 in: $U$ && 0&239 & 0&187 & 0&239 & 0&294 && 0&161 & 0&157 & 0&211 & 0&281 && 0&197 & 0&252 & 0&309 & 0&325 \\
 out: $U$ && 0&208 & 0&077 & 0&208 & 0&230 && 0&021 & $-$0&003 & 0&057 & 0&169 && 0&050 & 0&163 & 0&299 & 0&264 \\
 in: KSS && 0&479 & 0&375 & 0&478 & 0&587 && 0&321 & 0&313 & 0&423 & 0&562 && 0&394 & 0&504 & 0&618 & 0&650 \\
 out: KSS && 0&416 & 0&155 & 0&416 & 0&459 && 0&042 & $-$0&005 & 0&113 & 0&339 && 0&099 & 0&325 & 0&599 & 0&529 \\
\hline 
\multicolumn{28}{l}{Panel D: in-sample calibrating period (1885 -- 1999) / out-of-sample predicting period (2000 -- 2003) } \\
 in: $A$ && 0&245 & 0&350 & 0&248 & 0&140 && 0&414 & 0&370 & 0&317 & 0&185 && 0&233 & 0&181 & 0&172 & 0&105 \\
 out: $A$ && 0&522 & 0&690 & 0&523 & 0&330 && 0&768 & 0&702 & 0&725 & 0&461 && 0&539 & 0&463 & 0&418 & 0&278 \\
 in: $D$ && 0&709 & 0&707 & 0&712 & 0&725 && 0&725 & 0&669 & 0&726 & 0&734 && 0&733 & 0&669 & 0&773 & 0&747 \\
 out: $D$ && 0&731 & 0&821 & 0&745 & 0&533 && 0&883 & 0&820 & 0&855 & 0&769 && 0&784 & 0&664 & 0&764 & 0&700 \\
 in: $U$ && 0&232 & 0&178 & 0&232 & 0&292 && 0&156 & 0&149 & 0&204 & 0&275 && 0&250 & 0&244 & 0&301 & 0&321 \\
 out: $U$ && 0&104 & 0&065 & 0&111 & 0&102 && 0&058 & 0&059 & 0&065 & 0&154 && 0&122 & 0&100 & 0&173 & 0&211 \\
 in: KSS && 0&463 & 0&357 & 0&463 & 0&585 && 0&311 & 0&298 & 0&409 & 0&549 && 0&500 & 0&487 & 0&601 & 0&642 \\
 out: KSS && 0&209 & 0&130 & 0&223 & 0&204 && 0&115 & 0&118 & 0&130 & 0&308 && 0&244 & 0&201 & 0&345 & 0&422 \\
\hline 
\multicolumn{28}{l}{Panel E: in-sample calibrating period (1885 -- 2006) / out-of-sample predicting period (2007 -- 2009) } \\
 in: $A$ && 0&248 & 0&351 & 0&249 & 0&144 && 0&414 & 0&367 & 0&315 & 0&186 && 0&236 & 0&258 & 0&172 & 0&107 \\
 out: $A$ && 0&660 & 0&711 & 0&666 & 0&333 && 0&633 & 0&603 & 0&640 & 0&423 && 0&557 & 0&616 & 0&471 & 0&175 \\
 in: $D$ && 0&710 & 0&708 & 0&713 & 0&715 && 0&733 & 0&678 & 0&734 & 0&730 && 0&736 & 0&746 & 0&774 & 0&742 \\
 out: $D$ && 0&855 & 0&902 & 0&875 & 0&900 && 0&913 & 0&869 & 0&887 & 0&933 && 0&892 & 0&911 & 0&863 & 0&969 \\
 in: $U$ && 0&231 & 0&179 & 0&232 & 0&286 && 0&160 & 0&155 & 0&209 & 0&272 && 0&250 & 0&244 & 0&301 & 0&317 \\
 out: $U$ && 0&097 & 0&095 & 0&104 & 0&283 && 0&140 & 0&133 & 0&124 & 0&255 && 0&168 & 0&147 & 0&196 & {\bf 0}& {\bf 397} \\
 in: KSS && 0&462 & 0&357 & 0&463 & 0&571 && 0&319 & 0&311 & 0&419 & 0&544 && 0&500 & 0&489 & 0&603 & 0&634 \\
 out: KSS && 0&195 & 0&191 & 0&209 & 0&567 && 0&279 & 0&266 & 0&247 & 0&510 && 0&335 & 0&294 & 0&392 & {\bf 0}& {\bf 793} \\
\hline 
\multicolumn{28}{l}{Panel F: in-sample calibrating period (1885 -- 2010) / out-of-sample predicting period (2011 -- 2015) } \\
 in: $A$ && 0&258 & 0&350 & 0&254 & 0&140 && 0&326 & 0&439 & 0&312 & 0&183 && 0&230 & 0&256 & 0&169 & 0&100 \\
 out: $A$ && 0&180 & 0&352 & 0&174 & 0&128 && 0&229 & 0&358 & 0&229 & 0&136 && 0&155 & 0&195 & 0&112 & 0&066 \\
 in: $D$ && 0&729 & 0&712 & 0&725 & 0&726 && 0&747 & 0&752 & 0&737 & 0&738 && 0&740 & 0&751 & 0&780 & 0&758 \\
 out: $D$ && 0&667 & 0&636 & 0&682 & 0&667 && 0&682 & 0&696 & 0&682 & 0&714 && 0&737 & 0&720 & 0&842 & 0&571 \\
 in: $U$ && 0&235 & 0&181 & 0&235 & 0&293 && 0&211 & 0&157 & 0&213 & 0&277 && 0&255 & 0&248 & 0&306 & 0&329 \\
 out: $U$ && {\bf 0}&{\bf 243} & 0&142 & {\bf 0}&{\bf 254} & 0&269 && {\bf 0}&{\bf 227} & {\bf 0}&{\bf 169} & {\bf 0}&{\bf 227} & {\bf 0}&{\bf 289} && {\bf 0}&{\bf 291} &{\bf 0}&{\bf 262} & {\bf 0}&{\bf 365} & 0&253 \\
 in: KSS && 0&470 & 0&362 & 0&471 & 0&587 && 0&421 & 0&313 & 0&426 & 0&554 && 0&510 & 0&495 & 0&611 & 0&658 \\
 out: KSS && {\bf 0}&{\bf 486} & 0&285 & {\bf 0}&{\bf 508} & 0&538 && {\bf 0}&{\bf 453} & {\bf 0}&{\bf 338} & {\bf 0}&{\bf 453} & {\bf 0}&{\bf 578} && {\bf 0}&{\bf 582} &{\bf 0}&{\bf 525} & {\bf 0}&{\bf 730} & 0&505 \\

\bottomrule
\end{tabular}
\end{table*}

\section{Conclusion}

We have performed a recurrence interval analysis of financial extremes
in the DJIA index during the period from 1885 to 2015. We determine the extreme returns 
according to a newly proposed extreme identifying approach, as well as quantile thresholds. 
With the extreme identifying approach we are able to locate the optimal extreme threshold associated with the minimum KS statistics
of tail distributions. We find that the recurrence intervals, which are
the periods of time between the successive extremes of different types
of returns and thresholds, follow a $q$-exponential distribution.  This
allows us to analytically derive the hazard probability $W(\Delta t |
t)$ that within the time interval $\Delta t$ since the last extreme
event that occurred at time $t$ we will observe the next extreme
event. The analytical $W(\Delta t | t)$ value agrees well with the
empirical hazard probability estimated from real data.

Using the hazard probability, we develop an extreme-return-prediction
model for forecasting imminent financial extreme events. When the hazard
probability is greater than the hazard threshold, this model can warn
when an extreme event is about to occur. The hazard threshold is
obtained by maximizing the usefulness of extreme forecasts. Both
in-sample tests and out-of-sample predictions reveal that the signals
generated by our prediction model are better statistically than the
benchmark of neglecting these signals and that the input distribution
formula used to fit the recurrence intervals has no influence on the
final outcome of our early warning model. Although in most cases the
predictive performance of in-sample tests are better than that of
out-of-sample predictions, expanding the in-sample calibrating period
could yield out-of-sample predictions that are better than in-sample
tests. In addition, increasing the extreme-extracting threshold could
improve the predictive power of our model in both in-sample tests and
out-of-sample predictions. Our results may shed new light the occurrence
of extremes in financial markets and on the application of recurrence
interval analysis to forecasting financial extremes.

\section*{Acknowledgments}

Z.-Q.J. and W.-X.Z. acknowledge support from the National Natural
Science Foundation of China (71131007 and 71532009), Shanghai ``Chen
Guang'' Project (2012CG34), Program for Changjiang Scholars and
Innovative Research Team in University (IRT1028), China Scholarship
Council (201406745014) and the Fundamental Research Funds for the
Central Universities. G.-J.W. and C.X. acknowledge support from the
National Natural Science Foundation of China (71501066, 71373072, and
71521061). A.C. acknowledges the support from Brazilian agencies FAPEAL
(PPP 20110902-011-0025-0069/60030-733/2011) and CNPq (PDE
20736012014-6). H.E.S. was supported by NSF (Grants CMMI 1125290, PHY
1505000, and CHE- 1213217) and by DOE Contract (DE-AC07-05Id14517).

\section*{Reference}
\bibliographystyle{elsarticle-harv}

\end{document}